\documentclass[a4paper,fleqn,usenatbib]{mnras}

\usepackage[T1]{fontenc}
\usepackage{ae,aecompl}
\usepackage{upgreek}

\usepackage{graphicx}	
\usepackage{amsmath}	
\usepackage{physics}
\usepackage{amssymb,bm}	
\usepackage{xspace}	
\def\app#1#2{%
  \mathrel{%
    \setbox0=\hbox{$#1\sim$}%
    \setbox2=\hbox{%
      \rlap{\hbox{$#1\propto$}}%
      \lower1.1\ht0\box0%
    }%
    \raise0.25\ht2\box2%
  }%
}
\def\approxprop{\mathpalette\app\relax}

\def\Sref#1{Sect.~\ref{#1}\xspace}
\def\Fref#1{Fig.~\ref{#1}\xspace}
\def\Tref#1{Table~\ref{#1}\xspace}
\def\Eref#1{Eq.~\ref{#1}\xspace}
\def\Aref#1{Appendix~\ref{#1}\xspace}

\graphicspath{{./images/}}

\usepackage[table,xcdraw,svgnames]{xcolor}

\newcommand{\corr}[1]{\textcolor{black}{#1}}

\usepackage{newtxtext,newtxmath}

\title[GL of GWs: Effect of Microlens Population]{Gravitational Lensing of 
Gravitational Waves: Effect of Microlens Population in Lensing Galaxies
}

\author[Anuj Mishra et al.]{Anuj Mishra,$^{1}$\thanks{E-mail: anuj@iucaa.in}
Ashish Kumar Meena,$^{2}$
Anupreeta More,$^{1,3}$\thanks{E-mail: anupreeta@iucaa.in}
Sukanta Bose$^{1,4}$ and Jasjeet Singh Bagla$^{2}$
\\
\\
$^{1}$The Inter-University Centre for Astronomy and Astrophysics (IUCAA), 
Post Bag 4, Ganeshkhind, Pune 411007, India
\\
$^{2}$Indian Institute of Science Education and Research (IISER) Mohali,
Knowledge City, Sector 81, Sahibzada Ajit Singh Nagar, Punjab 140306,
India
\\
$^{3}$Kavli Institute for the Physics and Mathematics of the Universe (IPMU), 5-1-5 Kashiwanoha, Kashiwa-shi, Chiba 277-8583, Japan
\\
$^{4}$Department of Physics \& Astronomy, Washington State University, Pullman, WA 99164, USA
}

\date{Accepted XXX. Received YYY; in original form ZZZ}

\pubyear{2021}

\begin{document}
\label{firstpage}
\pagerange{\pageref{firstpage}--\pageref{lastpage}}
\maketitle

\begin{abstract}
\corr{With increasing sensitivities of the current ground-based gravitational-wave (GW) detectors, the prospects of detecting a strongly lensed GW signal are going to be high in the coming years. When such a signal passes through an intervening lensing galaxy or galaxy cluster, the embedded stellar-mass microlenses lead to interference patterns in the signal that may leave observable signatures. 
In this work, we present an extensive study of these wave effects in the LIGO/Virgo frequency band ($10$-$10^4$ Hz) due to the presence of the microlens population in galaxy-scale lenses for the first time.
We consider a wide range of strong lensing (macro) magnifications and the corresponding surface microlens densities found in lensing galaxies and use them to generate realisations of the amplification factor.
The methodologies for simulating amplification curves for both types of images (minima and saddle points) are also discussed.
We then study how microlensing is broadly affected by the parameters like macro-magnifications, stellar densities, the initial mass function (IMF), types of images, and microlens distribution around the source.
In general, with increasing macro-magnification values, the effects of microlensing become increasingly significant regardless of other parameters. 
Mismatch analysis between the lensed and the unlensed GW waveforms from chirping binaries suggests that, while inferring the source parameters, microlensing can not be neglected for macro-magnification $\gtrsim 15$. 
Furthermore, for extremely high macro-magnifications $\gtrsim 100$, the mismatch can even exceed $5\%$,
which can result in both a missed detection and, consequently, 
a missed lensed signal.
}
 
\end{abstract}


\section{Introduction} 
\label{sec:intro}

The detection of gravitational waves \citep[GWs, e.g.,][]{2019PhRvX...9c1040A,
2020arXiv201014527A} by the Laser Interferometer Gravitational-wave Observatory (LIGO) and the Virgo opened up a new window to observe the Universe. 
Those observatories have so far announced the detection of 50 GW events in their first
three observing runs, and this number will continue to rise with upcoming observations 
and detector facilities, like the Kamioka Gravitational Wave Detector~\citep[KAGRA,][]{2012CQGra..29l4007S}.
These observed GW events are the results of merging black holes (BH-BH), neutron stars 
(NS-NS), and black hole-neutron star (BH-NS) binaries in galaxies at cosmological distances. 

Since these are cosmologically distant sources, the possibility of gravitational lensing 
of their GW signals is quite promising.
On the theoretical side, gravitational lensing of 
GWs by galaxy- and cluster-scale lenses and its applications have been 
investigated in several works ~\citep[e.g.,][]{2017NatCo...8.1148L, 2017ApJ...835..103T, 
2018MNRAS.476.2220L, 2018IAUS..338...98S, 2018arXiv180205273B, 2019arXiv190103190B,
2020arXiv200613219B}. 
Searches have also been carried out for signatures of
strong lensing in LIGO and Virgo data~\citep[e.g.,][]{2019ApJ...874L...2H, 
2019MNRAS.485.5180S, 2020arXiv200712709D, 2020arXiv200906539L}.
Since the wavelength of GWs in the LIGO band is much smaller compared to the 
Schwarzschild radius of a galaxy or a galaxy cluster$-$scale lens, one can use the usual geometrical optics approach to calculate the strong lensing 
effects~\citep[e.g.,][]{2017arXiv170204724D, 2020arXiv200712709D,2020arXiv200812814M,2018arXiv180707062H}. 
Strong lensing amplifies the GW signal (independent of frequency) by a factor 
$\sqrt{\mu}$, where $\mu$ is the strong lensing magnification, and hence, increases 
the signal-to-noise ratio (SNR) by the same factor.
The amplitude of the GW signal is inversely proportional to the luminosity 
distance to the source and proportional to the chirp mass of the source binary. 
Hence, strong lensing introduces degeneracies in the measurement of these two 
parameters and can lead to misleading results~\citep[e.g.,][]{2018arXiv180205273B,
2019arXiv190103190B, 2020arXiv200613219B}. 

However, this is not the only effect for lensing of GWs in the LIGO frequency range as the GW wavelength is of the order of the Schwarzschild 
radius of \corr{compact} objects with mass range $\sim 10$~M$_\odot-10^4$~M$_\odot$
(e.g., Figure 1 in~\citealt{2020MNRAS.492.1127M}).
As a result, the diffraction effects become essential and introduce 
frequency-dependent features in the GW signal \citep[e.g.][]{1999PhRvD..59h3001B, 1999PThPS.133..137N, 2003ApJ...595.1039T, PhysRevLett.122.041103}.
Due to the frequency dependence, the effect of microlensing is not only limited to 
the luminosity distance or chirp mass, instead, it could affect most of the GW parameters.
In fact, \citet{2018PhRvD..98j3022C} showed that the effect of microlenses 
having mass $\gtrsim 30$~M$_\odot$ can be detected with the current ground-based detectors 
provided that SNR $\gtrsim 30$. 
\citet{2020MNRAS.492.1127M} discussed the effect of point mass microlenses combined 
with strong lensing on the GW signal and pointed out the possibility that 
microlensing can affect two counterparts of the strongly lensed signal in entirely 
different ways leading to misidentification of the strongly lensed GW signals. 

\corr{Most of these aforementioned microlensing studies are focused on the isolated point mass lens or a point mass lens in the presence of external effects. 
However,}
in a realistic gravitational lensing scenario, each of the strongly lensed GW
signals will be affected by a population of microlenses. 
In such cases, a macroimage will split into several microimages which will interfere and form complicated interference patterns 
\citep[see][for a python based pipeline to find microimages]{2020A&A...643A.167P}. 
The first study in this regard was done by~\citealt[][henceforth referred to as D19]{2019A&A...627A.130D}, 
in which the authors discuss the possibility of microlensing due to a population of 
microlenses present in a galaxy halo or the intra$-$cluster medium, on the strongly 
lensed GW signals.
In D19 and their subsequent work \citep{2020PhRvD.101l3512D}, the microlensing effect 
is considered for surface stellar density $\approx12$~M$_\odot\:\textrm{pc}^{-2}$ and 
for very high strong lensing magnifications only, where caustics overlap and microlensing 
effects become inevitable.
However, these specific conditions may not hold for a majority of strongly lensed 
signals as the local surface densities and strong lensing magnifications for the 
lensed counterparts can vary significantly.   

\corr{In this work,} we look at the effect of 
microlens populations on different  counterparts
of
strongly lensed GW signals. 
Following~\cite{2019MNRAS.483.5583V}, we consider \corr{a wide range of} typical strong-lensing magnifications in combination with typical stellar densities in galaxy$-$scale lenses to determine the microlensing effects. 
To estimate the amplification due to a microlens population, we have developed a code in Mathematica following the method described in \citet[][henceforth, referred to as UG95]{1995ApJ...442...67U}. 
\corr{We study the effect for both minima and saddle points macroimages, and the methodologies for simulating their amplification curves are also discussed.
We then study how microlensing is broadly affected by the parameters like macro-magnifications, stellar densities, the IMF, types of images, and microlens distribution around the macroimage of the source. 
To quantify the effect of microlensing on GW signals (from coalescing binaries), we compute the GW \textit{match} between the unlensed and the lensed signals. Such an analysis helps us estimate the effect on the detectability of GWs and in the inference of the GW source parameters, i.e., on the \textit{effectualness} and \textit{faithfulness} of the GW signals \citep{1998PhRvD..57..885D}, respectively.
To our knowledge, this is the first detailed study of how various parameters related to a galaxy-lens govern the wave-effects in the LIGO/Virgo frequency range.}

A few terms will be used throughout the paper. We refer to the galaxy lenses as macrolenses. 
The embedded stars and stellar remnants in these macrolenses that perturb them on the small scale are referred to as microlenses, and their surface density is denoted by $\Sigma_\bullet$. 
The images produced by the macrolens are referred to as macroimages 
while their magnification is referred to as macro-magnification ($\mu$). Since we measure 
signal amplitude in the case of GWs rather than flux, it gets amplified by a factor 
$\sqrt{\mu}$ in such cases, which we refer to as macro-amplification value. 
Around the positions of the microlenses, several images of a macroimage can form, 
which we refer to as microimages. 
For our analysis, we assume the redshifts of the lens and the source to be 
$z_{\rm d}=0.5$ and $z_{\rm s}=2.0$, respectively.

This paper is organized as follows. In \Sref{sec:basic_lensing}, we describe the 
basics of gravitational lensing in the geometric and wave optics regimes relevant for 
this work. In \Sref{sec:magnification factor}, we describe the formalism and various 
numerical tests which validate our methodology of computing amplification factors in the 
case of microlens population. The methodologies for simulating amplification curves for 
both type-I (minima) and type-II (saddle) images are also discussed.
In \Sref{sec:microlens population}, we mention the generation and simulation of a 
microlens population embedded in a galaxy$-$scale strong lens.
In \Sref{sec:res_disc}, we present the results and discuss the importance of various 
factors such as macro$-$magnification, surface microlens densities, and stellar IMF. 
The conclusions are summarized in \Sref{sec:conclusions}.

\section{Basic Theory of Gravitational Lensing}
\label{sec:basic_lensing}
In this section, we briefly discuss the basics of gravitational lensing 
in geometric~\citep{1992grle.book.....S} and wave optics limits~\citep{2003ApJ...595.1039T}.
In the geometric optics limit, the gravitational lensing of a source at an  angular diameter 
distance $D_{\rm s}$ due to the presence of an intervening lens/deflector at an angular 
diameter distance $D_{\rm d}$ can be described by the so-called gravitational lens equation 
(assuming small-angle and thin-lens approximation),
\begin{equation}
    \pmb{y} = \pmb{x}-\pmb{\alpha}(\pmb{x})
    = \pmb{x}-\nabla_{\hspace{-0.05cm}\pmb{x}}\psi(\pmb{x}),
    \label{eq:lens equation}
\end{equation}
where $\pmb{y} =\pmb{\eta}D_{\rm d}/(\xi_0 D_{\rm s})\equiv\pmb{\beta}/\theta_0$ and 
$\pmb{x} = \pmb{\xi}/\xi_0\equiv\pmb{\theta}/\theta_0$ represent the projected unlensed 
source position and the lensed image position on the lens/image plane (measured with 
respect to the optical axis), respectively. 
Here $\pmb{\eta}=\pmb{\beta}D_{\rm s}$ and $\pmb{\xi}=\pmb{\theta}D_{\rm d}$ represent 
physical distances on the source and image planes, respectively,
while $\pmb{\beta}$ and $\pmb{\theta}$ are their corresponding angular positions on the sky. 
\corr{In order to make the lens equation dimensionless, we have chosen an arbitrary length scale $\xi_0$ (or, angular scale $\theta_0$) such that $\xi_0\equiv\theta_0 D_{\rm d}$. }
The lens equation simply describes 
vector addition and is derived purely from geometry where physics is contained 
in the deflection term $\pmb{\alpha}\left(\pmb{x}\right)$, i.e., in the projected 2D lensing 
potential $\psi\left(\pmb{x}\right)$, which determines the deflection as a function of the 
impact parameter $\pmb{x}$ on the lens plane. The nonlinearity brought by the deflection 
term is what leads to the formation of multiple images of a given source. In the context 
of GWs, this would lead to multiple detection of the same source which may be separated in 
the time domain by an order of a few minutes to several years (e.g., see fig. 13 in 
\citealt{2018MNRAS.480.3842O}). The magnification factor corresponding to these different 
lensed macroimages (or, equivalently, events) is given as
\begin{equation}
    \mu\equiv |\det\mathbb{A}|^{-1} = \big[\left(1-\kappa\right)^2-\gamma^2\big]^{-1},
    \label{eq:macro mag}
\end{equation}
where $\mathbb{A}_{ij}\equiv[\partial y_i/\partial x_j]$ is the Jacobian (matrix) corresponding 
to the lens equation~\Eref{eq:lens equation} while $\kappa$ and $\gamma$ represent the convergence and shear at the image position, respectively. Both $\kappa$ and $\gamma$ are functions of the lens plane coordinate $\pmb{x}\equiv (x_1,x_2)$.

For a given lensed signal, the corresponding time delay with respect to its unlensed counterpart is given by
\begin{equation}
    t_{\rm d}\left(\pmb{x},\pmb{y}\right) = T_{\rm s}
    \left[\frac{1}{2}|\pmb{x}-\pmb{y}|^2 - \psi\left(\pmb{x}\right)
    + \phi_{\rm m}\left(y\right)\right]\equiv T_{\rm s} \tau_{\rm d}(\pmb{x},\pmb{y})\,,
    \label{eq:general time delay}
\end{equation}
\corr{where $\phi_m\left(y\right)$ is a constant independent of lens properties and the factor $T_{\rm s}$ is the characteristic time delay defined via  
\begin{equation}
    \frac{T_{\rm s}}{\left(1+z_{\rm d}\right)}=\xi_0^2\frac{D_{\rm s}}{cD_{\rm d} D_{\rm ds}} 
     \equiv \frac{2{R_{\rm s}}_0}{c}=\frac{4GM_0}{c^3}\,.
     \label{eq: T_s}
\end{equation}
Above, $z_{\rm d}$ is the lens redshift, $c$ is the speed of light, $D_{\rm ds}$ is the angular 
diameter distance between the source and the lens,  and ${R_{\rm s}}_0$ is the Schwarzschild radius corresponding to the mass $M_0$. This mass has an Einstein radius $\xi_0$ if placed on the 
lens plane.  The factor $T_{\rm s}$ roughly sets the order of the time delay for a given lens system.}

The formalism of geometrical/ray optics described above is valid as long as the time delay 
between any two images is sufficiently large compared to the wavelength $\lambda$ of light, 
i.e., $f t_{\rm d}\gg 1$, where $f$ is the frequency of the signal. This relation holds in 
a typical scenario of strong gravitational lensing where different macroimages are formed. 
Gravitational lensing of gravitational waves due to galaxies or galaxy cluster scale lenses 
can also be described using the above formalism.
However, if the time delay of a lensed signal is less than or of the order of its time period, 
i.e., when $f t_{\rm d}\lesssim 1$ (or, equivalently, ${R_{\rm s}}_0\lesssim \lambda$), 
then wave effects are non-negligible and one has to take diffraction into account. 
Furthermore, when the source and the deflector are far from the observer, one can use the Huygens-Fresnel principle for analyzing the lensing of the incoming plane-wave flux, 
in which case every point on the lens plane acts as a secondary source (Huygens point sources),
and the amplitude of the signal at each point on the observer plane is the superposition of 
the signals from these various sources, leading to interference patterns. 

For an isolated point mass lens of mass $M_{\rm L}$, the above condition ($f t_{\rm d}\lesssim 1$)
translates to, roughly, $M_{\rm L}\lesssim 10^5M_\odot(f/$Hz$)^{-1}$.
Therefore, for gravitational waves with frequency in the LIGO band ($10-10^4$~Hz), the mass range where wave effects become significant is $\sim 10-10^4$~M$_\odot$. 
This mass range is predominantly responsible for microlensing in the strongly lensed images 
of a source. In comparison, for electromagnetic (EM) signals with $f\sim10^6-10^{20}$~Hz, the 
diffraction effects become significant for \corr{the} mass range $\sim10^{-15}-10^{-1}M_\odot$. This is 
a major difference between the microlensing of EM waves and that of GWs.

In \corr{the case} of microlensing, one has to consider the corrections arising from 
wave optics \citep[e.g.,][]{1999PThPS.133..137N, 2003ApJ...595.1039T}. If we denote the ratio 
of the observed lensed and the unlensed GW amplitudes as $ F\left(f,\pmb{y}\right)$, then the 
amplification of the lensed signal is given by the diffraction integral 
\citep{1992grle.book.....S, goodman2005introduction}
\begin{equation}
    F\left(\nu,\ \pmb{y}\right) = \frac{\nu}{i}
    \int d^2\pmb{x} \: {\rm exp}\left[2 \pi i \nu \tau_{\rm d}\left(\pmb{x},\pmb{y}\right)\right],
    \label{eq:general amp fac}
\end{equation}
where
\begin{equation}
    \tau_{\rm d}(\pmb{x},\pmb{y})=\frac{t_{\rm d}(\pmb{x},\pmb{y})}{T_{\rm s}},\ \ 
    \nu\equiv \frac{\xi_0^2D_{\rm s}}{D_{\rm d}D_{\rm ds}}\frac{f}{c}(1+z_{\rm d})=T_{\rm s} f.
\end{equation} 
Note that the definition of dimensionless frequency, $\nu$, and dimensionless time, $\tau_{\rm d}$, 
is such that $\nu \tau_{\rm d}=f t_{\rm d}$. Since $F(f)$ is a complex-valued function, the total 
amplification, $|F|$, and phase shift, $\theta_F$, can be obtained through the relation $F(f)=|F|e^{i\theta_F}$.
As is apparent, in wave optics the amplification is frequency-dependent, unlike 
in geometric optics, where the average magnification over a frequency range is independent of the frequency.
In the geometric optics limit ($f \gg t_{\rm d}^{-1}$), the integral in \Eref{eq:general amp fac} 
becomes highly oscillatory and only the stationary points of the time-delay surface contribute to 
the amplification. In that case, wave optics reduces to ray optics and, as a result, the diffraction 
integral reduces to
\begin{equation}
    F\left(f\right)\big|_{\rm geo} = \sum_j \sqrt{|\mu_j|} \: 
    {\rm exp}\left(i 2 \pi f t_{d,j} - i \pi n_j\right),
    \label{eq:general amp fac geo}
\end{equation}
where $\mu_j$ and $t_{d,j}$ are, respectively, the magnification factor and the time 
delay for the $j$-th image. Also, $n_j$ is the Morse index, with values 0, 1/2, and 1 for 
stationary points corresponding to minima, saddle points and maxima of the time-delay surface, 
respectively.
As one can see from \corr{the} above equation, even in the geometric optics limit, gravitational
lensing introduces an extra phase -- the so-called Morse phase -- of $e^{-i\pi/2}$ and 
$e^{-i\pi}$ in the saddle points and maxima with respect to the minima, 
respectively (\citealt{2017arXiv170204724D}, \citealt{ 2020arXiv200812814M}). 
This phase difference can be used to search for the strongly lensed and multiply imaged 
gravitational wave signals, and to constrain viable lenses \citep{2020arXiv200712709D}.

The diffraction integral, \Eref{eq:general amp fac}, can be solved analytically only 
for some trivial lens models.
For example, the solution for a point mass lens of mass $M_{\rm L}$ is given by
\begin{equation}
    \begin{split}
        F\left(\omega,y\right) = \exp\bigg\{\frac{\pi \omega}{4} + 
        \frac{i\omega}{2}\left[\ln\left(\frac{\omega}{2}\right) - 
        2\phi_{\rm m}\left(y\right)\right]\bigg\} \\
        \times\ \Gamma\left(1-\frac{i\omega}{2}\right)
        {}_{1}{F}_{1}\left(\frac{i\omega}{2},1;\frac{i\omega y^2}{2}\right),
    \end{split}
    \label{eq:amp fac point}
\end{equation}
where $\omega = 8\pi G(1+z_{\rm d})M_{\rm L} f/c^3\equiv 2\pi \nu$, 
$\phi_{\rm m}(y) = (x_{\rm m} - y)^2/2 - \ln(x_{\rm m})$
and $x_{\rm m} = \left(y+\sqrt{y^2+4}\right)/2$. 
The scale radius, $\xi_0$, has been chosen equivalent to the Einstein radius of the lens. 
In the geometric optics limit ( $f \gg t_{\rm d}^{-1}$), the above equation can be written as
\begin{equation}
    F\left(f\right)\big|_{\rm geo} = \sqrt{|\mu_{+}|} - i\sqrt{|\mu_{-}|} 
    \exp\left(2\pi i f \Delta t_{\rm d} \right),
    \label{eq:amp fac point geo}
\end{equation}
where $\mu_{+}$ and $\mu_{-}$ are the amplification factors for primary and secondary
images formed due to a point mass lens, and $\Delta t_{\rm d}$ is the time delay 
between these two images.

However, in \corr{a realistic scenario of strong lensing}, the possibility of a GW signal encountering a 
massive isolated point lens is very less relative to it encountering a microlens population. \corr{Typically}, microlensing of strongly lensed images happens 
due to the population of point mass lenses instead of a single point mass lens. 
As a result, the time delay factor, $t_{\rm d}$, in \Eref{eq:general time delay} is 
modified and includes a contribution from the macromodel in terms of the convergence 
and shear at the image position, and the population of microlenses near the macroimage.
\corr{The resultant potential then becomes $\psi\rightarrow \psi_{\rm total} = \psi_{\text{\scriptsize SL}} + \psi_{\text{\scriptsize ML}}$, where $\psi_{\text{\scriptsize SL}}$ is the macromodel 
potential and $\psi_{\text{\scriptsize ML}}$ is the lens potential due to the microlensing population embedded in the macromodel.} 
These contributions are given by~\citep{mollerach2002gravitational, 2011MNRAS.411.1671S}
\begin{equation}
    \begin{split}
        \psi_{\text{\scriptsize ML}}\left(\pmb{x}\right)& = \sum_{\rm k}\frac{m_{\rm k}}{M_0}\ln|\pmb{x}-\pmb{x}_{\rm k}|, \\
        \psi_{\text{\scriptsize SL}}\left(\pmb{x}\right)  & = 
        \frac{\kappa}{2}\left(x_1^2+x_2^2 \right) +
        \frac{\gamma_{1}}{2}\left(x_1^2-x_2^2 \right) + 
        \gamma_{2}x_1 x_2,
    \end{split}
    \label{eq:potential pop}
\end{equation}
where $m_{\rm k}$ and $\pmb{x}_{\rm k}$ denote, 
respectively, the mass and position of the $k$-th point mass lens in the population; $M_0$ 
is an arbitrary mass value as defined in \Eref{eq: T_s};
$\kappa$ and $\left(\gamma_1,\gamma_2\right)$ represent, respectively, the convergence and 
the components of shear introduced due to the presence of the macrolens. 
\corr{Here, we assumed constant $\kappa$ and $\gamma$ values since they are slowly varying. }
The diffraction integral, \Eref{eq:general amp fac}, with the above lens potential, containing 
population of microlenses, cannot be solved analytically, in which case one has to use numerical
methods, such as in UG95 and D19, to obtain an approximate solution.

\section{Calculation of the Amplification Factor}
\label{sec:magnification factor}
Except for the most trivial lens models, like that of an isolated point mass, the 
potential $\psi(\pmb{x})$ takes a complicated form, in which case no analytical 
form can be derived straightforwardly. Furthermore, it is highly inefficient to 
numerically integrate the diffraction integral, \Eref{eq:general amp fac}, because 
of the oscillatory nature of the integrand, and the fact that the direct calculation 
of $F(f)$ is a three-dimensional problem in $x_1$, $x_2$ and $f$. Hence, one needs 
to use a numerical method that is more efficient and resolves the problems mentioned 
above. Such numerical methods have been described in UG95 and D19. 
In the current work, we follow the method of UG95 to calculate the magnification 
factor that is described below in \Sref{ssec:formalism}. Subsequently, we also 
demonstrate the validity of our code for both \corr{minima and saddle points} macroimages. In Sects. \ref{ssec: testing code_pnt_lens} and \ref{ssec: testing code_type-I_pnt+shear}, 
we consider microlensing for two elementary cases, namely, an isolated 
point mass lens and a point lens situated near a \corr{minima-type} macroimage in the presence of an external shear. 
Generally, simulating amplification curves for \corr{saddle points} is nontrivial, and 
we discuss the issue separately in \Sref{ssec: testing code_saddle_type-II}. We also 
describe our methodology to deal with saddle points macroimages and perform numerical 
tests to verify our results for simple lensing configurations.

\subsection{Formalism}
\label{ssec:formalism}
By using the methods of contour integration and Fourier transformation ($\mathcal{F}$), 
UG95 splits the problem of calculating $F(f)$ into two parts and reduces it into two 
dimensions as described below. Firstly, we define $\Upsilon(\nu)\equiv iF(\nu)/\nu$\corr{, assuming $\pmb{y}$ to be fixed. 
Then, we have
\begin{equation}
    \begin{split}
        \mathcal{F}[\Upsilon(\nu)] \equiv  \widetilde{F}(\tau ') = & \int {\rm d}^2\pmb{x}
        \int d\nu \exp \left(i2\pi \nu [\tau_{\rm d}(\pmb{x})-\tau ']\right) \\
    \Rightarrow\ \    \widetilde{F}(\tau ') = &\int {\rm d}^2\pmb{x} ~\delta[\tau_{\rm d}(\pmb{x})-\tau '],   
    \end{split}
    \label{eq:F(t) general}
\end{equation}
}
where $\tau ' \equiv t/T_{\rm s}.\ $Now, using $\nu=T_{\rm s}f$ and the fact that
$\mathcal{F}^{-1}[\widetilde{F}(\tau ')]=\Upsilon(\nu)$, we get
\begin{equation}  
   F(f)=\frac{f}{i}\int {\rm d}t \exp\left(i2\pi f t\right) \widetilde{F}(t),
   \label{eq:F(f) ulmer}
\end{equation}
where `$t$' represents the time delay value relative to an arbitrary reference time. 
For a \corr{minima-type} macroimage, it is usually measured relative to the global minima 
of the time delay surface, which marks the arrival of the first microimage. Whereas for 
macroimages at \corr{saddle points}, we measure `$t$' relative to the arrival of the dominant 
saddle image (discussed in \Sref{ssec: testing code_saddle_type-II}).
\corr{Also, since we will have a finite range of $\widetilde{F}(t)$ values in an actual computation, we would need to further use an apodization function in \Eref{eq:F(f) ulmer} that removes the erroneous contribution from the edges. Otherwise, the computed values for both $|F|$ and $\theta_F$ would be significantly inaccurate and will show oscillatory behaviour at lower frequencies. In our analysis, we have used a cosine window function (e.g., see D19) that removes these irregularities and produces an excellent output, as discussed in the next subsections.} 

Equation~\ref{eq:F(f) ulmer} can then be evaluated as a contour integral. 
The area between the curves defined by $\tau_{\rm d}(\pmb{x},\pmb{y})=\tau '$ and
$\tau_{\rm d}(\pmb{x},\pmb{y})=\tau '+d\tau '$ is $A=\widetilde{F}(\tau ')d\tau '$ up to first order. 
This area can also be evaluated as an integral $A=\oint ds dl$, where $ds$ is 
the infinitesimal length along the contour and 
$dl=d\tau '/|\nabla_{\hspace{-0.05cm}\pmb{x}}\tau_{\rm d}|$ is the orthogonal distance between 
the two contours at the point of evaluation. 
Moreover, there can in general be more than one such contour. 
Thus, \corr{by comparison of the areas evaluated using these two methods}, we finally get
\begin{equation}
   \widetilde{F}(\tau ')=\sum_{\rm k} \oint_{C_{\rm k}}
   \frac{ds}{|\nabla_{\hspace{-0.05cm}\pmb{x}}\tau_{\rm d}|}.
   \label{eq:F(t) contour}
\end{equation} 
The summation is over all the contours, \corr{$C_{\rm k}$}, where $\tau_{\rm d}(\pmb{x},\pmb{y})=\tau '$. 
Thus, for a given time-delay function (or lensing potential), we first 
compute $\widetilde{F}(t)$ using \Eref{eq:F(t) contour} and 
then inverse Fourier transform it back to get the required $F(f)$, as in
\Eref{eq:F(f) ulmer}. 
Also, from \Eref{eq:F(t) contour}, one can see that $\widetilde{F}(t)$ is a smooth 
function except at critical time $t_i$ where the images form, i.e., where 
$|\nabla_{\hspace{-0.05cm}\pmb{x}}\tau_{\rm d}|=0$. 
The reader is referred to the appendix of UG95 for the method to handle these singularities.

\subsection{Testing Numerical Code: Isolated Point lens}
\label{ssec: testing code_pnt_lens}

\begin{figure*}
    \centering
    \includegraphics[scale=0.22]{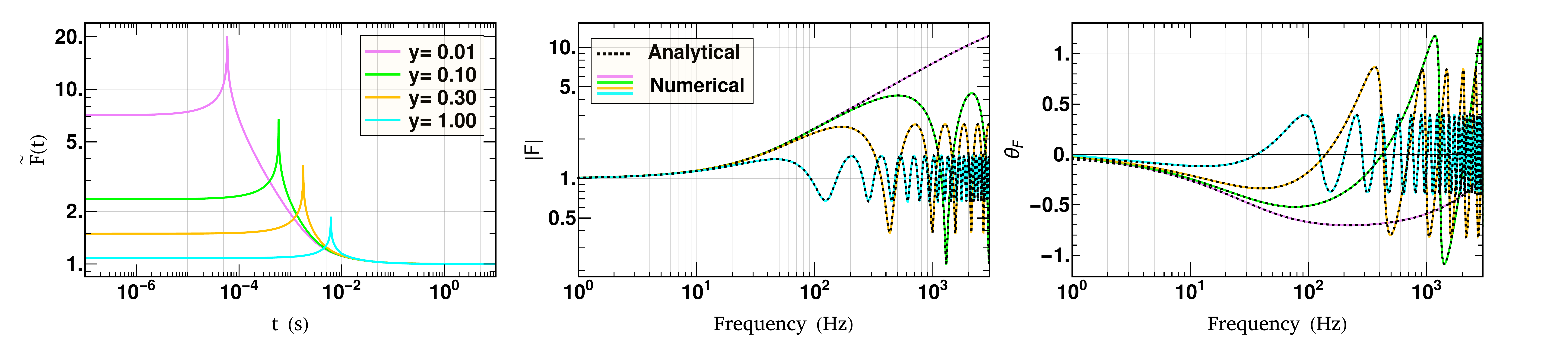}
    \vspace{-0.7cm}
    \caption{\corr{Test of numerical code shown for an isolated point lens of $100$~M$_\odot$ at $z_{\rm d}=0.5$ for a source at $z_{\rm s}=2$. The analysis is done for four different values of the impact parameter $y=\beta/\theta_0=\{0.01,~0.1,~0.3,~1.0\}$. 
    \textit{Left:} The curves show numerically computed $\widetilde{F}(t)$ normalised by a factor of $2\pi$ so that its value approaches unity in the no$-$lens limit (large time delays). 
    \textit{Middle and Right:} The curves show the comparison between the analytical and the numerically computed frequency-dependent amplification factor $F(f)=|F|e^{i\theta_F}$. The solid coloured curves have been numerically obtained from the $\widetilde{F}(t)$ curves using \Eref{eq:F(f) ulmer}, whereas the dotted black curves denote analytical results (see \Eref{eq:amp fac point}).}
    }
    \label{fig: NTest - point mass lens}
\end{figure*}

Since we have the analytic form of $F(f)$ for an isolated point mass 
lens, \Eref{eq:amp fac point}, it can be used as an initial testing 
ground for our numerical code based on the above formalism. Hence, in this subsection,
we compare $F(f)$ generated via two independent methods: analytical and numerical.
We consider a $100$~M$_\odot$ point mass lens placed at a lens redshift $z_{\rm d}=0.5$,
and a gravitational wave source placed at $z_{\rm s}=2$.
The analysis has been done for four different non-zero source positions:
$y=\beta/\theta_0\in\{0.01,~0.1,~0.3,~1.0\}$.
In the case of a point mass lens, two images of opposite parities are always formed, 
where positive and negative parities correspond to the minimum and \corr{the saddle point} of the time-delay surface, respectively. The magnification of each image and the (dimensionless) time delay between 
them is given by
\begin{equation}
    \mu_\pm=\frac{1}{2}\pm\frac{y^2+2}{2y\sqrt{y^2+4}},\ \Delta\tau_{\rm d}=\frac{y\sqrt{y^2+4}}{2}+\ln{\left(\frac{\sqrt{y^2+4}+y}{\sqrt{y^2+4}-y}\right)}\,.
    \label{eq:mag_td_pnt_lens}
\end{equation}

In the left panel of \Fref{fig: NTest - point mass lens}, we show the normalised $\widetilde{F}(t)$ curves, computed using \Eref{eq:F(t) contour}, for different source positions. The x-axis represents the time delay measured with respect to 
the global minimum (situated at $t=0$).
Since we are interested in the LIGO frequency range of $10-10^4$ Hz, we need to compute $\widetilde{F}(t)$ within the range $\sim 10^{-6}- 1\rm~$sec such that we cover the region where $ft_{\rm d}\lesssim 1$.
So, \corr{we  generate $\widetilde{F}(t)$} values \corr{for a sufficient number of time-points}
within this interval and  interpolate \corr{between them} using the Hermite interpolation method to obtain a continuous function $\widetilde{F}(t)$, which can be easily inverse Fourier transformed to obtain $F(f)$ using~\Eref{eq:F(f) ulmer}.
The $\widetilde{F}(t)$ curves shown are normalised such that their value approaches one in the no lens limit (large time delays). 
Since the time delay surface approaches a paraboloid at large values, this normalization is done by dividing the obtained curves 
by $2\pi$ (since \Eref{eq:F(t) contour} yields 2$\pi$ in the no lens limit, i.e., for 
circular contours). In this way, the curves start from a value corresponding to the 
amplification of the image formed at the minimum, i.e., $\sqrt{\mu_+}$, and eventually 
approach unity at large time delay. Between these two expected behaviours, it encounters a 
logarithmic divergence corresponding to the saddle point (image with negative parity). 
The time delay at the point of this divergence is the time delay between the two images 
(since the first image occurs at $t=0$).  As expected, we can see that this time delay 
between the images decreases as we move towards the lens (eventually becoming zero when 
$y=0$) while the amplitude of the logarithmic pulse increases, increasing the magnification 
of the saddle point image. 

In the middle and right panels of \Fref{fig: NTest - point mass lens}, we show the comparison 
between the analytical results (obtained using \Eref{eq:amp fac point}) and the numerical 
results for the computation of the amplification factor $F(f)$=$|F|\exp(i\theta_F)$. 
The black-dotted lines represent the $|F(f)|$ and $\theta_F$ calculated 
using the analytical formula given by 
\Eref{eq:amp fac point}, respectively. 
The different solid-coloured lines represent the $|F(f)|$ and $\theta_F$ values, which have 
been computed numerically using our code. For the numerical computation, we first generate 
$\widetilde{F}(t)$ values using \Eref{eq:F(t) contour} and substitute it in \Eref{eq:F(f) ulmer} along with a \corr{cosine window function} (because of the finite range of $\widetilde{F}(t)$). \corr{As previously mentioned, without this apodization the computed $F(f)$ values will show oscillatory behaviour, especially below 100 Hz.} 

As one can see from \Fref{fig: NTest - point mass lens}, the agreement between analytical 
and numerical values is excellent. In all cases, the factor $|F|$ approaches unity and the 
phase factor $\theta_F$ approaches zero as we go lower in the frequency ($f\ll t^{-1}$), 
which means that the lens is invisible for signals with large wavelengths compared to the 
Schwarzschild radius of the lens ($\lambda \gg {R_{\rm s}}_0$). 
The wave effects start to appear when $\lambda \sim {R_{\rm s}}_0$ ($ft\sim 1$), which causes 
modulation in the amplification factor. 
As the frequency increases ($\lambda \ll {R_{\rm s}}_0$), wave optics approaches ray optics, 
\corr{i.e., $F(f)$ oscillates rapidly about its geometric optics limit}
and the average magnification over a frequency range becomes independent of the frequency (as in strong lensing). 
In the frequency range shown in the figure, only the blue curve ($y=1.0$) has been able to approach the ray optics limit approximately. 
In general, for a point lens, the average values of $|F|$ and $\theta_F$ at high frequencies approach $\sqrt{\mu_+}$ and zero, respectively, in accordance with \Eref{eq:amp fac point geo}.
We notice that our numerical code recovers all the features mentioned for $\widetilde{F}(t)$ and $F(f)$ very well. 

\subsection{Testing Numerical Code: Type-I (Minima) Macroimages}
\label{ssec: testing code_type-I_pnt+shear}
\begin{figure*}
    \centering
    \includegraphics[scale=0.22]{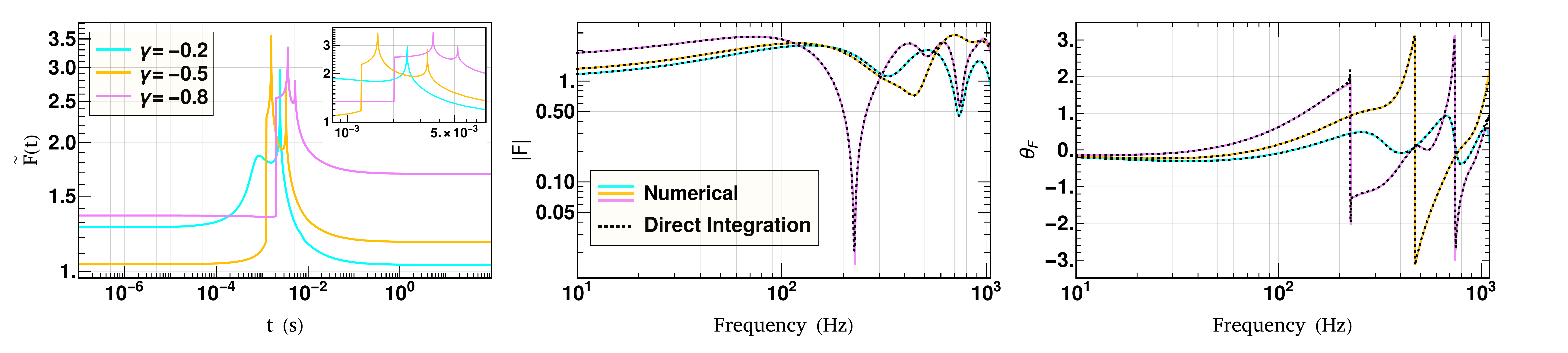}
    \vspace{-0.7cm}
    \caption{
    \corr{Test of numerical code shown for a point mass microlens of $100$~M$_\odot$ at $z_{\rm d}=0.5$ in the presence of shear. The source is kept at $z_{\rm s}=2$ and the source position is fixed to  $(y_1,y_2)=0.4(\cos(\pi/8),\sin(\pi/8))$. 
    The analysis is done for three different values of the shear $\gamma=\{-0.2,-0.5,-0.8\}$. 
    \textit{Left:} The curves show numerically computed $\widetilde{F}(t)$ (using \Eref{eq:F(t) contour}) normalised by a factor of $2\pi$, which ensures that they approach their no-microlens (strong lensing) limit, $\sqrt{\mu}=(1-\gamma^2)^{-1/2}$, at large time-delay values. 
    \textit{Middle and Right:} The curves show the comparison between the numerical and the direct evaluation methods for computing the frequency-dependent amplification factor $F(f)=|F|e^{i\theta_F}$. The solid coloured curves have been numerically obtained from the $\widetilde{F}(t)$ curves using \Eref{eq:F(f) ulmer}, while the dotted black curves are obtained via direct (numerical) integration of \Eref{eq:general amp fac}. }
    }
    \label{fig: NTest - point mass with external effects}
\end{figure*}

In this subsection, we test our code for a slightly complicated case where
we place a point lens of $100{\rm M}_\odot$ close to a \corr{minima-type} macroimage \corr{of a source} in the presence of an external shear
($\gamma$) with no convergence ($\kappa=0$). 
\corr{Without loss of generality, one can always choose the principle direction of the shear to be horizontally aligned, in which case $\gamma_2=0$, $|\gamma|=\sqrt{\gamma_1^2+\gamma_2^2}=|\gamma_1|$. Also, we place the macroimage at the origin of the source plane coordinates, i.e., at $\pmb{y}=(0,0)$. Unless stated otherwise, we adopt this reference frame throughout our analysis.}
\corr{Now, in this case,} the effective lens potential in 
\Eref{eq:general time delay} can be written as
\begin{equation}
    \psi_{\rm total} = \ln\left(\sqrt{x_1^2+x_2^2}\right) + 
    \frac{\gamma}{2}\left(x_1^2-x_2^2 \right). 
    \label{eq:external effect potential}
\end{equation}
For the potential written above, we do not have an analytic solution for the diffraction integral, 
unlike in the case of \corr{a} point mass lens. Therefore, to perform the numerical test for this case, we 
directly evaluate the double integral in \Eref{eq:general amp fac} numerically and compare it with 
the one obtained via our code. However, the direct evaluation is slow and does not work well for 
higher frequencies where the integrand becomes too oscillatory.     

For the computation of the amplification factor, the comparison between the direct numerical 
evaluation (dotted-black lines) and the one obtained via our code (solid-coloured lines) is 
shown in the middle and right panel of \Fref{fig: NTest - point mass with external effects}. 
The analysis has been done for three different values of shear, $\gamma$=$\{-0.2$, $-0.5$, $-0.8\}$, 
keeping the source position fixed at $(y_1,y_2)=0.4(\cos(\pi/8),\sin(\pi/8))$. Again, we observe 
that there is excellent agreement between direct numerical integration and the adopted numerical 
method of UG95.
Also, the amplification curves approach the strong lensing amplification value for low frequency 
values (no$-$microlensing limit) and the phase shift curves approach zero, as expected. 

The corresponding calculations of $\widetilde{F}(t)$ are shown in the left panel of 
\Fref{fig: NTest - point mass with external effects}. We have again normalised the plots by dividing 
the originally obtained ones by a factor of $2\pi$. This normalization ensures that the curves 
approach to value $\sqrt{\mu}=(1-\gamma^2)^{-1/2}$ in their no$-$microlensing (strong lensing) 
limit at large time delays.
The time delay function, in this case, includes four stationary points, at least two of which 
are always real. 
In \corr{the} $\widetilde{F}(t)$ plots in \Fref{fig: NTest - point mass with external effects}, we observe 
that two microimages form in the case of $\gamma=-0.2$ (blue curve). One of these corresponds to 
the global minima (discontinuity at $t=0$) and other is for the saddle point (logarithmic peak 
at $t\sim 2.5$~ms). The other two values of the shear lead to a four-microimage geometry (orange and magenta curves). These microimages correspond to the two minima at low time delay (two 
discontinuities) and the two saddle points at a higher time delay (two logarithmic peaks).

The direct evaluation of the diffraction integral (\Eref{eq:general amp fac}), adopted here for 
comparison with our code, cannot be used in the case of a microlens population (or for any 
nontrivial potential), as it becomes highly inefficient and does not perform well at higher 
frequencies because of the oscillatory integrand.
Hence, this method can not be used further and we solely rely on the method by UG95 to compute 
the amplification factor $F(f)$, using our code, throughout our analysis in the paper.

\subsection{Testing Numerical Code: Type-II (Saddle) Macroimages}
\label{ssec: testing code_saddle_type-II}
In this subsection, we describe the numerical scheme that is used to compute the amplification factor 
for a \corr{saddle point} image. Unlike in the case of minima, here \corr{the time delay} contours neither 
close locally nor have \corr{a} global minima, as they are hyperbolic in nature rather than elliptical. 
However, if one chooses a sufficiently large region, the contribution from the neighborhood of a 
\corr{saddle point} is given by
\begin{equation}
    \widetilde{F}(t)= -2\sqrt{|\mu_-|}\ \log{|t-t_i|} + \text{non-singular part + constant}
\end{equation}
where $t_i$ and $\mu_-$ denote the time delay and magnification value corresponding to the 
saddle point, respectively, and the constant depends on the size of the region. By sufficiently 
large, we mean the size of the region should be such that $|u^{-2}\mu_-|^{-1/2}\gg 1$ near the 
boundary, where $u$ denotes the arc parameter of the contour (the reader is referred to Appendix 
B of UG95 for further details). The presence of a constant does not affect the computation of 
$F(f)$, especially when the integration range is chosen carefully, such that 
$\Re{ \int {\rm d}t\ e^{i 2\pi f t}} = 0$, and at higher frequencies where 
$\Im{ \int {\rm d} t\ e^{i 2\pi f t}} \ll 1$.  

When a saddle point macroimage splits into microimages, there will always be a dominant saddle microimage 
which will dominate $\widetilde{F}(t)$ and $F(f)$. In our simulation, we first find this image and 
measure the time delay values relative to this image, i.e., we fix the arrival time of the dominant 
saddle image at $t=0$ (this arbitrary value is chosen for simplicity). Given a time delay function 
corresponding to a saddle point macroimage, one can find the location of the dominant saddle image numerically 
by iteratively computing the minima along the coordinate direction with increasing curvature and maxima 
along the coordinate direction with decreasing curvature. 
We then compute $\widetilde{F}(t)$ values symmetrically about $t=0$ and inverse Fourier transform 
it to get the required amplification factor values $F(f)$. A time range of $\mathcal{O}(2)\rm~s$ 
is sufficient for most cases since only the region closer to the divergence would mainly contribute. 
This is due to the fact that the contribution of the nearly flat part of $\widetilde{F}(t)$ in the 
inverse Fourier transform will mostly be averaged out. 

\begin{figure*}
    \centering
    \includegraphics[scale=0.22]{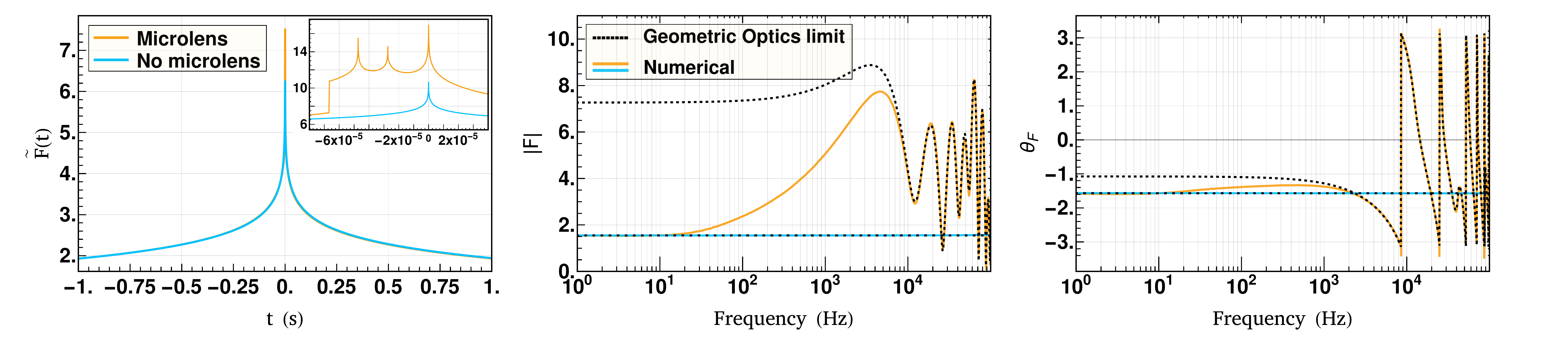}
    \vspace{-0.7cm}
    \caption{\corr{Test of our numerical code for the case of microlensing of a saddle point macroimage of a source at $z_{\rm s}=2$. The analysis is shown for two cases, with and without the presence of a 100 M$_\odot$ microlens at  $z_{\rm d}=0.5$.
   \textit{Left:} The curves show numerically computed $\widetilde{F}(t)$ obtained using \Eref{eq:F(t) contour}. The presence of microlens, in this case, leads to a four-microimage configuration (see orange curve in the inset of the left-most panel).
   \textit{Middle and Right:} The curves show the comparison between the numerical and the analytical computation of $F(f)=|F|e^{i\theta_F}$ in the geometrical optics limit ($ft_{\rm d}\gg 1$). The solid coloured curves have been numerically obtained from the $\widetilde{F}(t)$ curves using \Eref{eq:F(f) ulmer}, while the dotted black curves are obtained using \Eref{eq:general amp fac geo}. For the no microlens case, the geometric optics limit is equivalent to the strong lens limit ($ft_{\rm d}\ll 1$), while in general, the geometrical optics limit is reached at high frequencies $\gtrsim 10^4$ Hz where the $F(f)$ and $F(f)\big|_{\rm geo}$ match (see dotted black and orange curves).} 
   }
   \label{fig: Ntest - saddle test}
\end{figure*}

We test our numerical recipe for two cases, namely, in the absence of microlens and in the presence 
of a $100$ M$_\odot$ microlens leading to a four$-$microimage configuration. For both cases, we fix 
our macro-magnification value to $\mu=-2.4$. In the case of no microlens, one expects to recover 
strong lensing values for the amplification factor, i.e., $|F|=\sqrt{|\mu|}$ and $\theta_F=-\pi/2$, 
since there will be no interference in the absence of microlenses. We indeed recover these values 
as shown in the middle and right panel of \Fref{fig: Ntest - saddle test}, where $|F|=\sqrt{2.4}$ 
and $\theta_F=-\pi/2$ (the Morse phase shift of saddle-type macroimages). 
The dotted black curve represents the geometrical optics limit, which in this case is equivalent to 
the strong lensing value, while the blue curve represents the $F(f)$ values as obtained through the code numerically.

In the presence of a $100$~M$_\odot$ microlens, we keep the source inside the caustic to get a 
four$-$microimage geometry.  The time delay and magnification of these microimages are 
$\left(\frac{t}{10^{-5}{\rm s}},~\mu\right)\in\ \{(-6.68,11.98),~(-4.73,-3.76),~(2.73,-2.63),~(0.,-8.00)\}$.
These microimages correspond to the discontinuity and spikes in the $\widetilde{F}(t)$ as shown 
in the left panel of \Fref{fig: Ntest - saddle test}. Using \Eref{eq:general amp fac geo}, one 
can then find the $F(f)$ in the geometrical optics limit $(ft_{\rm d}\gg 1)$ and then compare 
it with the computed $F(f)$ at high frequencies. The comparison is shown in the middle and right 
panel of \Fref{fig: Ntest - saddle test}, where the dotted black curve represents the geometrical 
optics limit of $F(f)$ obtained using \Eref{eq:general amp fac geo} and 
\corr{the} solid orange curve shows the numerically computed $F(f)$ using the $\widetilde{F}(t)$ curve as shown in \Fref{fig: Ntest - saddle test}.

As we can see, for both cases, the numerically computed $F(f)$ and the expected geometrical 
optics limit values are in excellent agreement. Furthermore, in \Fref{fig: Ntest - saddle test}, 
the $\widetilde{F}(t)$ curves in both cases contain a dominant logarithmic divergence at $t=0$ 
(the arrival time of the dominant saddle image) and is, roughly, symmetric about it, as expected. 
The difference in $F(f)$ below $\lesssim 10^4~$Hz is mainly due to the diffraction effects and 
clearly demonstrates why one needs to incorporate wave optics in such cases.

\section{Microlens population embedded in a Macromodel}
\label{sec:microlens population}

\subsection{Methodology and Assumptions}
\label{ssec:methodology}

We first define our macromodel which is kept fixed throughout our analysis.
We consider an isolated elliptical galaxy, as a lens, at redshift $z_{\rm d}$=0.5, and \corr{model} the 
smooth matter fraction with a singular isothermal ellipsoid (SIE) density profile  
and a velocity dispersion ($\sigma_{\rm vd}$) of 230~km~s$^{-1}$ (taken as a rough mean 
$\sigma_{\rm vd}$ from lens sample of \citealt{2013ApJ...777...98S}).
Next, we calculate the surface density of \corr{the} compact objects at any given position in the macromodel. 
The density profile of the population of compact objects is modeled using 
the S\'ersic profile (see Equation 8 in \citealt{2019MNRAS.483.5583V}).
Next, to determine the mass function of the compact objects, we use the Chabrier IMF Function~\citep{2003PASP..115..763C, 2013MNRAS.429.1725M} with the mass range 0.01~M$_\odot$ to 200~M$_\odot$. 
\corr{Since within a time period of $\sim 5$ Gyr stars with masses $\gtrsim 1.2 {\rm M}_\odot$ will become remnants~\citep{Paxton_2010}, we make use of the initial-final mass relation from the Binary Population And Spectral Synthesis \citep[BPASS,][]{2017PASA...34...58E} to infer the final mass of the evolved stars, while the lower mass stars are kept unchanged.}
As a result, the final population of single objects \corr{consists} of stars and stellar remnants (e.g., white dwarfs, neutron stars and black holes).
The total fraction of the mass in the mass range $m\in$(0.01, 0.08)~M$_\odot$ is around 5$\%$.
This mass range predominantly affects the frequencies above the higher end of the
LIGO frequency range and the relative error due to the removal of this mass range is about
$\sim \mathcal{O}(1)\%$ in the $F(f)$ curve for typical strong lensing amplification values.
As a result, for computational efficiency, we remove the microlenses below 0.08~M$_\odot$ from our population.
\corr{Since galaxies also contain binary systems, we include them following \cite{2013ARA&A..51..269D}, where
the binary fraction in various mass ranges is estimated based on different (local) observations.
Our final mass distribution of the microlenses is confined to a range $m_{\bullet}\in (0.08,\sim 28)$ M$_\odot$, and the average mass of a microlens in our simulation is $\sim 0.44$ M$_{\odot}$.}
In other words, the number of microlenses present per solar mass is $\sim 2.28$, where we are treating each binary system as a single microlens.

Since \corr{minima and saddle points} are the most common types of lensed images seen in galaxy$-$scale lenses, 
we consider the microlensing effects for these two images in our analyses. 
For an SIE lens, the values of 
$\kappa$ and $\gamma$ due to the macrolens, at the position of lensed images, are equal to each other, i.e.,
$|\kappa|=|\gamma|$~\citep[e.g.,][]{2019MNRAS.483.5583V}.
\corr{We consider sufficiently wide range of $(\sqrt{\mu},\Sigma_\bullet)_i$, where `$i$' in the subscript denotes the image type, for both minima ($i= \rm m$) and saddle point ($i=\rm s$) macroimages. For minima (saddle points), we consider five (three) different cases of $(\sqrt{\mu},\Sigma_\bullet)_i$, which are listed in the \Tref{tab:stellar_density}.}

\begin{table}
  \centering
  \caption{\label{tab:stellar_density} Lens parameter values for minima and saddle points used in simulations. 
  The $\left(\kappa,\ \gamma\right)$ are the local convergence and shear values due to the (smooth) macrolens 
  mass distribution. The $\kappa^{\rm star}$ is the local convergence due to the mass in compact objects. The 
  $\mu$ represents macro$-$magnification, and $\Sigma_\bullet$ represents the surface microlens density.}
   \begin{tabular}{lccccr} 
    \hline
    $\kappa$ & $\gamma$ & $\kappa^{s\rm tar}$ & $\sqrt{\mu}$ &  $\Sigma_\bullet$ (M$_\odot$~pc$^{-2}$) \\ 
    \hline
    \multicolumn{5}{c}{Minima} \\
    \hline
    0.276  & 0.276 & 0.013 &  1.49 &   27   \\
    0.354  & 0.354 & 0.024 &  1.85 &   50   \\
    0.413  & 0.413 & 0.035 &  2.40 &   72   \\
    0.467  & 0.467 & 0.046 &  3.87 &   95   \\
    0.495  & 0.495 & 0.052 & 10.01 &  108  \\
    \hline
    \multicolumn{5}{c}{Saddle points} \\
    \hline
    0.504  & 0.504 & 0.054 & 11.05 &  113  \\
    0.548  & 0.548 & 0.065 &  3.21 &  135  \\
    0.722  & 0.722 & 0.115 &  1.50 &  239  \\
    \hline
  \end{tabular}
\end{table}

\subsection{Details of simulations}

For each $(\sqrt{\mu},\Sigma_\bullet)_i$, we first simulate a large box with an area of about $100$~pc$^2$.
We then populate this box with microlenses of surface mass density $\Sigma_\bullet$ such that they follow the evolved mass function. 
\corr{This region is then divided into 36 equal-sized patches having an area of $\sim 2.8$~pc$^2$ each. 
We then compute the amplification factor $F(f)$ assuming a macroimage of the source at the center of each patch, thereby generating 36 realisations corresponding to each $(\sqrt{\mu},\Sigma_\bullet)_i$.
As before, we assume the source redshift to be $z_{\rm s}=2.0$.}
Dividing the larger box of density $\Sigma_\bullet$ into smaller patches allows us to have density variations. 
As a result, the patches can have densities $\Sigma_\bullet \pm \sigma_{\rm d}$ where $\sigma_{\rm d}$ is the scatter.
To obtain the amplification curves, we first have to generate $\widetilde{F}(t)$ curves, as explained in 
\Sref{sec:magnification factor}. Calculating the $\widetilde{F}(t)$ is the most computationally expensive part 
of the whole simulation, and depends strongly on the number of microlenses (as it increases the number of microimages).
Additionally, the macro-magnification also affects the computational cost because with increasing values, we get 
larger and more complicated time$-$delay contour structures since the macro$-$magnification amplifies the microlensing 
effects.

As discussed in \Sref{sec:magnification factor}, once we generate a sufficiently large number of points 
for $\widetilde{F}(t)$, we then interpolate it using \corr{the} Hermite interpolation method to obtain the function 
$\widetilde{F}(t)$. By taking the inverse Fourier transform, we get the required amplification factor, 
$F(f)=|F(f)|e^{i\theta_F(f)}$, in a given frequency range. For cases where $\Sigma_\bullet \gtrsim 100$~M$_\odot$, 
we perform an optimisation where we remove microlenses $m_{\rm l}\lesssim 0.2$~M$_\odot$ that are present outside 
an area of $\sim 1.5$~pc$^2$ from the center. In this way, we increase the computational efficiency substantially 
at the cost of introducing only $\sim \mathcal{O}(1)\%$ relative error at low macro$-$magnifications. However, 
for high macro$-$magnifications ($\mu\gtrsim 50$), the errors increase significantly, especially, at higher 
frequencies ($\gtrsim 10^3$ Hz). 
Therefore, our results for high values of $(\sqrt{\mu},\Sigma_\bullet)_i$ may underestimate the effects of microlensing at such high frequencies.       

\begin{figure*}
    \centering
    \includegraphics[scale=0.21]{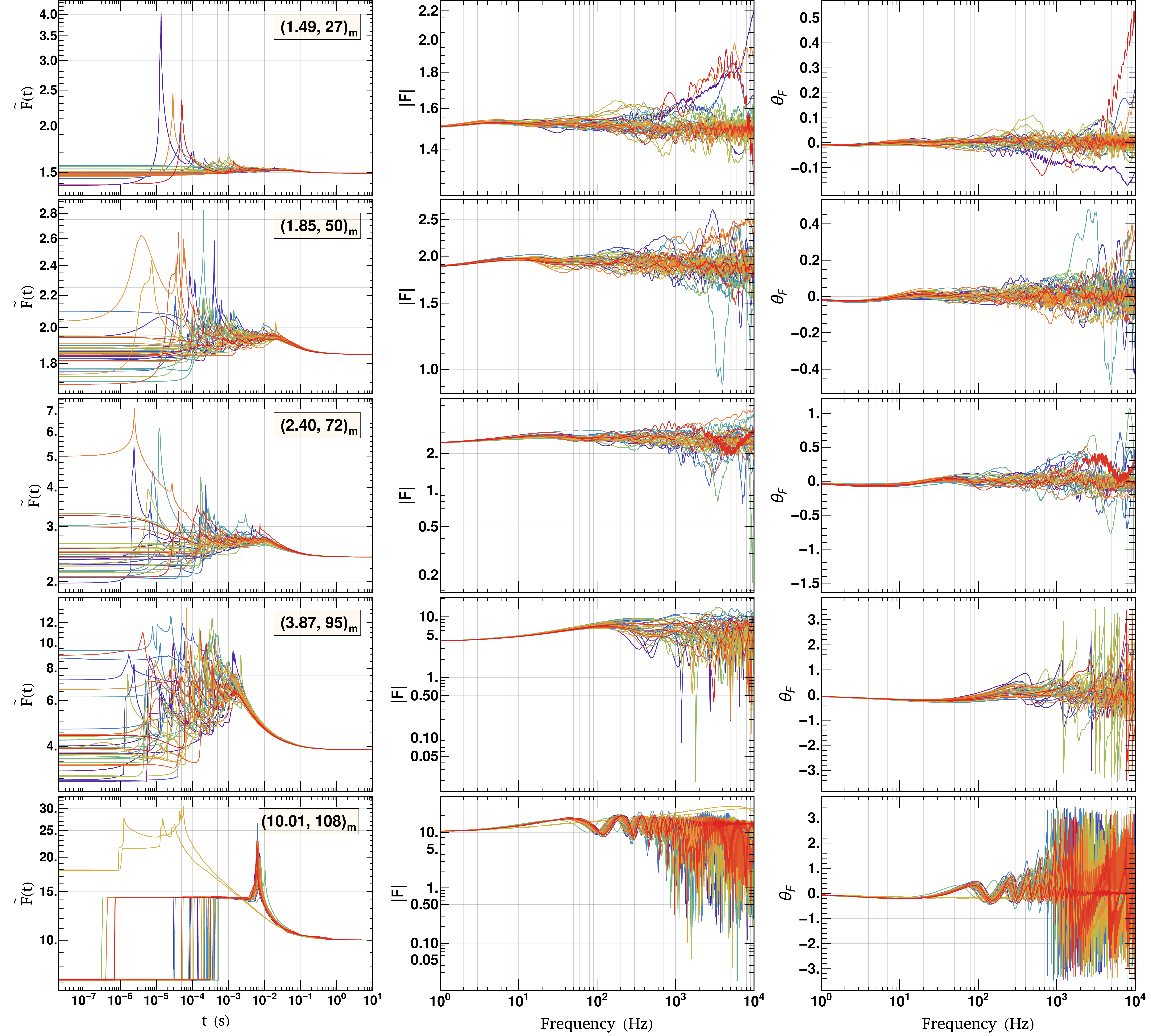}
    \caption{\corr{
     Effect of microlens population on the minima type macroimages. The analysis is done for five pairs of the macro amplification and the surface microlens density values as denoted by $(\sqrt{\mu},\Sigma_\bullet)_\mathrm{m}$ in the left-most panels. 
     These values are drawn from our SIE model (see~\Tref{tab:stellar_density}). \textit{Left}: Normalised $\widetilde{F}(t)$ curves, computed numerically using \Eref{eq:F(t) contour}. \textit{Middle and Right}:The corresponding amplification factor $F(f)=|F|e^{i\theta_F}$. Each row shows the analysis for all 36 realisations (coloured differently).}
        }     
    \label{fig:F(f) minima}
\end{figure*}

\begin{figure*}
    \centering
     \includegraphics[scale=0.31]{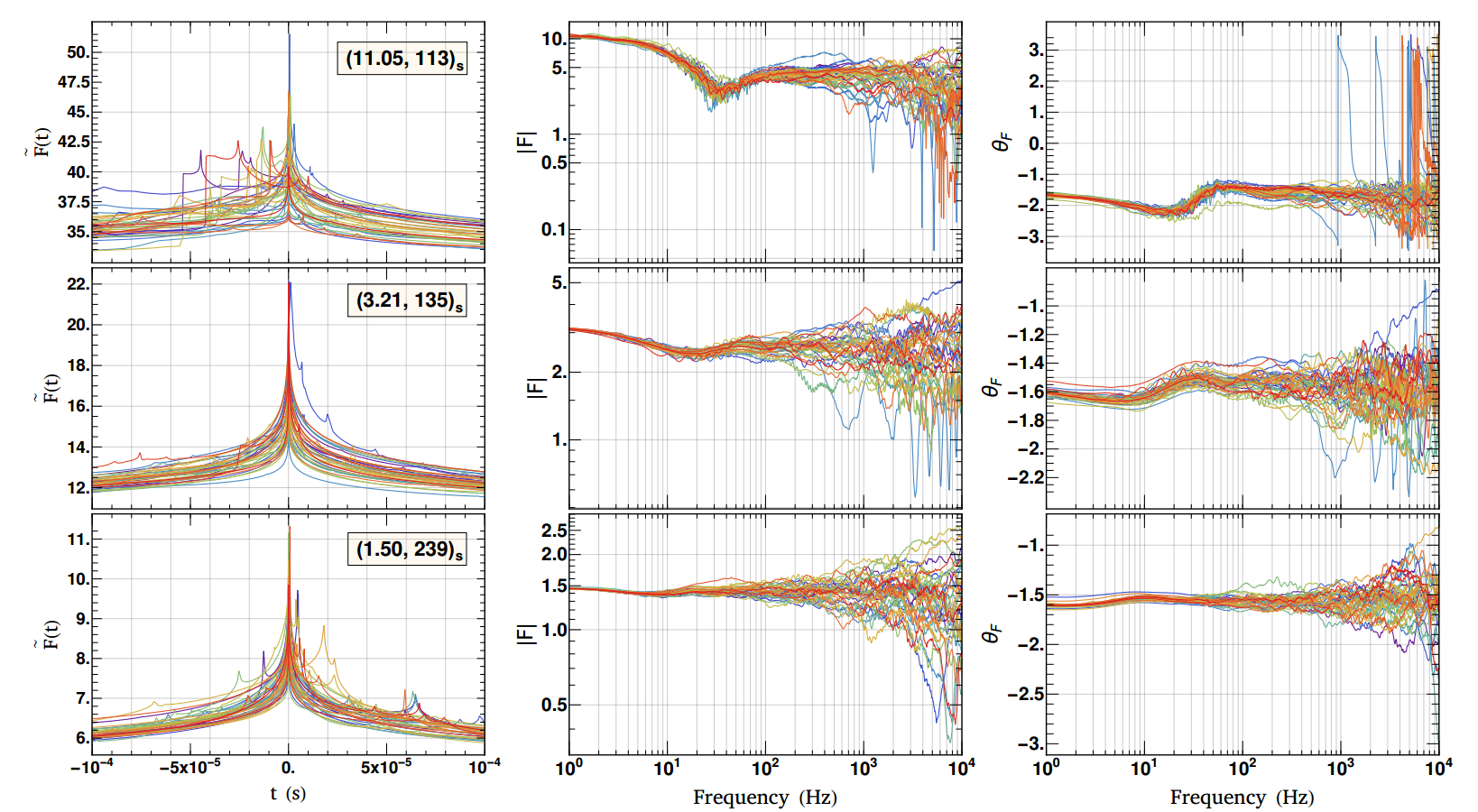}
    \caption{\corr{
      Effect of microlens population on the saddle point macroimages. The analysis is done for five pairs of the macro amplification and the surface microlens density values as denoted by $(\sqrt{\mu},\Sigma_\bullet)_\mathrm{s}$ in the left-most panels.
      These values are drawn from our SIE model (see~\Tref{tab:stellar_density}). \textit{Left}: Normalised $\widetilde{F}(t)$ curves, computed numerically using \Eref{eq:F(t) contour}. \textit{Middle and Right}: The corresponding amplification factor $F(f)=|F|e^{i\theta_F}$. Each row shows the analysis for all 36 realisations (coloured differently).}
     }
     \label{fig:F(f) saddle}
\end{figure*}

\section{Results and Discussion}
\label{sec:res_disc}
In this section, we present the results of our microlensing analysis. For this, we draw realisations of the amplification 
factor $F(f)$, using the formalism discussed in \Sref{sec:magnification factor}, for cases where a microlens population is embedded in a macromodel, as discussed in \Sref{sec:microlens population}. We then study the broad effects of microlensing 
on GW waveforms via mismatch analysis between the lensed and unlensed waveforms.

\subsection{Effect of Microlens Population: Type-I (Minima) Macroimages}
\label{ssec:ml_effect_minima}
As discussed in \Sref{sec:microlens population}, we generate realisations for five cases by varying the microlens 
surface densities, $\Sigma_\bullet$, and the macro-magnifications, $\mu$, found typically at the location of minima 
(see \Tref{tab:stellar_density}). 
Furthermore, we draw 36 realizations for each case to understand the typical and extreme situations, if any. 
The effects of microlens population on \corr{minima-type} images are shown in 
\Fref{fig:F(f) minima}. Each row shows all of the 36 realisations for each case. \corr{In each row, we show the $\widetilde{F}(t)$ curves, the absolute value of the amplification and the corresponding phase shift in the left, middle and right panels, respectively, due to the combined effect of strong lensing and microlensing.}

When inspecting any row, we find that, at low GW frequencies ($\sim$1~Hz), all \corr{$F(f)$} curves converge to the strong lensing 
amplification ($\sqrt{\mu}$) and to a phase shift of zero, since the oscillations due to microlensing are minimal to none. 
This behaviour is expected, as explained in \Sref{sec:basic_lensing} and \Sref{sec:magnification factor}. 
Towards higher frequencies, as the effects due to microlensing become significant, the \corr{$F(f)$} curves begin to show stronger 
oscillations. This is owing to the formation of many significant microimages with sufficiently long time-delays (such that $ft_{\rm d}\sim 1$) 
with respect to the macrominimum (global minima of the time delay surface).
Some realisations show extreme excursions from the rest of the set.
For example, \corr{the amplification curves of }the red and \corr{the purple} curve in the top-most \corr{row} show atypical behaviour because of the presence of a highly amplified microimage at low time delay value (see corresponding $\widetilde{F}(t)$ curves in the left panel). Such microimages in the low macro-magnification regime are highly dependent on the microlensing configuration in the neighbourhood of the source, which can conspire to mimic the effect of a heavy microlens placed closer to the source. We will explore this more robustly in \Sref{ssec:ml_effect_pmdist}.
From our analysis, we infer that our choice of drawing 36 realisations is reasonable enough to produce such extreme cases.

As we go to higher frequencies, the distortions increase and become more significant. Since the microimages from low 
mass microlenses have smaller time delay values, a greater number of microimages start contributing to the diffraction 
integral $F(f)$ at higher frequencies. As a result, the amplification factor $F(f)$ is more random and chaotic at 
higher frequencies for low values of macro-amplification ($\sqrt{\mu}$). However, as $\sqrt{\mu}$ increases, the amplification of microimages increases and can lead to the formation of multiple dominant microimages (microimages having high amplification) 
contributing to $F(f)$. Such contributions cause a relatively lesser chaotic but stronger modulations (see bottom two 
rows of \Fref{fig:F(f) minima}).  

Although not visible in the LIGO frequency range, at sufficiently high frequencies, we approach the geometrical optics 
regime where again the amplification becomes independent of frequency when averaged over a frequency range. The value of 
the magnification in the geometrical optics regime can be obtained, in principle, from the $\widetilde{F}(t)$ curves along 
with the rough estimate of the frequency after which this limit is reached. \corr{In the left panel of \Fref{fig:F(f) minima}}, the 
$\widetilde{F}(t)$ \corr{curves} at very low time delays ($\sim~10^{-7}$~s), where the curve is almost flat, gives the amplification at \corr{the} geometrical optics limit, as opposed to the macro-amplification at large $\widetilde{F}(t)$ (where all curves tend after $\sim~1$~s). 
Thus, observing the curves, we notice that the geometrical optics limit is reached at $\sim 10^6-10^7$~Hz 
with varying amplification values. This amplification at the geometric optics limit  tells us whether microlensing caused the 
overall amplification or deamplification of our signal. We notice that except for the case of $(10.01, 108)_\mathrm{m}$, \corr{more than} 
$50\%$ of the realisations had overall amplification. 
However, since our chosen area of the patch is not sufficient for the computation of the amplification factor at such 
high frequencies ($\gtrsim 10^5$ Hz), we have mostly underestimated our results in the geometrical optics limit. 
Therefore, in real case scenarios, we should expect a net amplification for most of the \corr{minima-type} images in the 
geometric optic limit, which is consistent with our observation in the LIGO frequency range.
The fact that minima tend to be amplified, on average, due to microlensing is consistent with microlensing studies in 
the electromagnetic (EM) domain \citep[e.g.,][]{2002ApJ...580..685S}. 

\begin{figure*}
    \centering
    \includegraphics[scale=0.352]{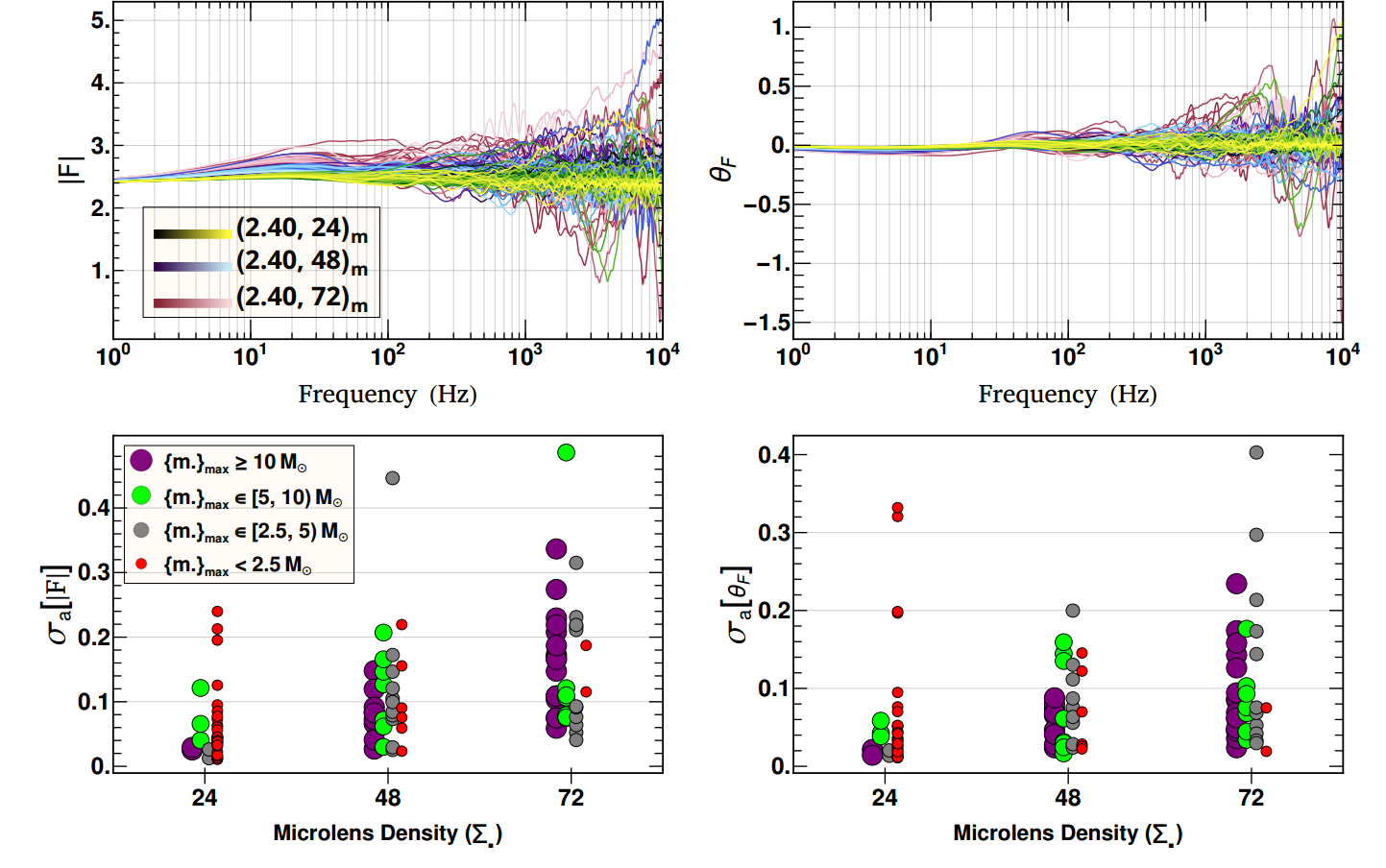}
    \caption{\corr{
    Effect of varying surface microlens density ($\Sigma_\bullet$). We show comparison between the amplification curves corresponding 
    to $\Sigma_\bullet\in \{24,\:48,\:72\}$~M$_\odot$ pc$^{-2}$, for a  fixed macro-amplification value of $\sqrt{\mu}=2.40$. \textit{Top}: the curves show the amplification factor $F(f)=|F|e^{i\theta_F}$ of 36 realisations for each $\Sigma_\bullet$.  \textit{Bottom}: we show the overall scatter due to microlensing and the effect of different microlens masses within the population. The $\sigma_{\text{\tiny a}}[|F|]$ and $\sigma_{\text{\tiny a}}[\theta_F]$ denote the normalised standard deviation of a realisation in $|F(f)|$ 
    and in $\theta_F(f)$ relative to the strong lensing values (no-microlens limit) in the LIGO frequency range 
    (see Eqs. \ref{eq: sigma_a_sdvar_analysis} and \ref{eq:sdv_G_a}). Here, each circle represents a specific realisation while its colour represents the mass range in which the maximum mass of that realisation, $\{m_\bullet\}_{\rm max}$, lies. A slight offset between the different coloured markers has been given to enhance visibility.}
    }
    \label{fig:sd_variation_analysis}
\end{figure*}

\subsection{Effect of Microlens Population: Type-II (Saddle point) Macroimages}
\label{ssec:ml_effect_saddle}
Contrary to minima, saddle point macroimages are formed at a relatively higher microlens densities and their 
macro-magnification values decrease with increasing microlens densities (see \Tref{tab:stellar_density}). Owing to 
this and the fact that the microlensing effects are not as significant for low macro-amplification values, we generate 
realisations for three cases only, thereby reducing the computational cost. As in the case of minima, every row of 
\Fref{fig:F(f) saddle} shows \corr{the $\widetilde{F}(t)$} curves, the absolute value of the amplification factor and the corresponding phase shifts in the left\corr{, middle} and right panels, respectively, for all of the 36 realisations. 
It is evident from \Fref{fig:F(f) saddle} that the microlensing effects are more 
significant at higher $\sqrt{\mu}$ rather than at high $\Sigma_\bullet$ value. 

For any \corr{row}, the \corr{amplification} curves converge to the macro-amplification value ($\sqrt{\mu}$) at lower frequencies and the scatter increases at higher 
frequencies similar to the case of minima. Moreover, we do recover the Morse phase in our simulations which causes an 
overall phase shift of $-\pi/2$ for \corr{saddle point} macroimages with respect to \corr{minima-type} macroimages 
(see \Eref{eq:general amp fac geo}).
Interestingly, in the top-most \corr{row} with the highest $\sqrt{\mu}$, we observe very strong modulations even at low
frequencies ($\sim 20-100$ Hz). As many of the GW signals tend to have a longer inspiral in the low-frequency regime, 
such a modulation at low frequencies is likely to affect the match filtering and parameter estimation significantly 
(see \Sref{ssec:ml_effect_mismatch}).
In general, we find saddle points tend to be deamplified on average, as opposed to minima, which is also consistent 
with microlensing studies in the EM domain.

Although the microlens densities of 113 ${\rm M}_\odot$~pc$^{-2}$ for saddle and 108 ${\rm M}_\odot$~pc$^{-2}$ 
for minima, and \corr{their} corresponding strong lens amplification ($\sqrt\mu$) \corr{values} are, roughly, similar to each other, the corresponding \corr{amplification factor} plots show significant differences (see $F(f)$ plots in the top-most row in \Fref{fig:F(f) saddle} and bottom-most row in \Fref{fig:F(f) minima}, respectively).
The overall modulations in case of saddle points $(11.05,113)_\mathrm{s}$ are stronger as compared to minima $(10.01,108)_\mathrm{m}$, especially at lower frequencies.

\subsection{Effect of varying surface microlens density}
\label{ssec:ml_effect_sd_var}

In this subsection, we discuss the effect of varying surface microlens density, $\Sigma_\bullet$, on the amplification 
factor $F(f)$, near both \corr{minima and saddle point} macroimages. We first discuss the effect of varying density near \corr{minima-type} macroimages, for which we fixed the macro-amplification to $\sqrt{\mu}=2.40$ and the comparison is done by drawing 36 realisations for each of the three sets of
$(\sqrt{\mu},\Sigma_\bullet)_i\in \{(2.40, 24)_\mathrm{m},\ (2.40, 48)_\mathrm{m},\ (2.40, 72)_\mathrm{m}\}$.
The results are shown in \Fref{fig:sd_variation_analysis}.

On average, the distortions and the scatter in the curves are proportional to the microlens density, i.e., the microlensing effects are most significant for the highest density considered. 
Also, as we increase the density, these distortions start 
becoming significant from relatively lower frequencies (see the top panel). 
Such behaviour is expected because, as the microlens density increases, it leads to the formation of a relatively greater number of significantly amplified microimages, especially at large time delays ($\sim 10^{-3}-10^{-2}$~s in the corresponding $\widetilde{F}(t)$ curves).
This can be further understood using \Eref{eq:potential pop}.
For instance, if there are $N$ number of microlenses of mass $m$ present in a small neighborhood around some $x_{\rm 0}$, i.e., if the position of the microlenses, $x_{\rm k}$, is such that $|x_{\rm k}-x_{\rm 0}|\ll 1$, then the net effect of those microlenses will be similar to that of a single microlens of mass $N m$ present at $x_{\rm 0}$. 

When studying the microlensing effect of a population, the other crucial factors apart from population density 
are macro-model potential in which it is embedded and the frequency range under consideration. As explained 
in \Sref{sec:basic_lensing}, the microlensing effect from a point lens of mass $m$ is significant at a frequency 
$f$ only when the order of the time delay between the two images $t_{\rm d}\gtrsim f^{-1}$. Due to this, in lower density populations where the abundance 
of heavy mass microlenses as well as density fluctuations is low, like in the case of $24$~M$_\odot$ pc$^{-2}$ in 
our population, the microlensing effects become significant at relatively higher frequencies $\gtrsim 10^3$ ~Hz. 

In the bottom row of  \Fref{fig:sd_variation_analysis}, we quantify the microlensing effect by computing the 
normalised standard deviation relative to the no$-$microlens limit ($|F|=\sqrt{\mu}$ and $\theta_{\rm F}=0$) for 
each realisation, i.e.,
\corr{
\begin{equation}
     \sigma_{\text{\scriptsize a}}^2=\frac{\int {\rm d}f ~G^{\text{\scriptsize a}}(f)}{\int {\rm d}f},
     \label{eq: sigma_a_sdvar_analysis}
\end{equation}
}
\corr{
such that,
\begin{equation}
      G^{\text{\scriptsize a}}(f)= 
\begin{cases}
    \left[\frac{|F(f)|}{\sqrt{\mu}}-1\right]^2,& \text{for}\hspace{0.2cm} \sigma_{\text{\scriptsize a}}[|F|]\\[8pt]
    \left[\theta_F(f)-0\right]^2,   & \text{for}\hspace{0.2cm} \sigma_{\text{\scriptsize a}}[\theta_F]
\end{cases}
    \label{eq:sdv_G_a}
\end{equation}
}
The median over the realisations corresponding to 
$\sigma_{\text{\scriptsize a}}[|F|]$ are, roughly, $4\%$, $8\%$ and $12\%$ for increasing density values, while for $\sigma_{\text{\scriptsize a}}[\theta_F]$ the values are $4\%$, $6\%$ and $8\%$.
Therefore, it is interesting to note that the standard deviation increases proportionally to the density values, showing that the overall scatter increases significantly with increasing density values.
\corr{Therefore, in  \Fref{fig:sd_variation_analysis},} the outliers present in the bottom row indeed correspond to the 
realizations that show atypical behaviour in the $F(f)$ curves as shown in the top row.
\begin{figure}
    \centering
    \includegraphics[scale=0.182]{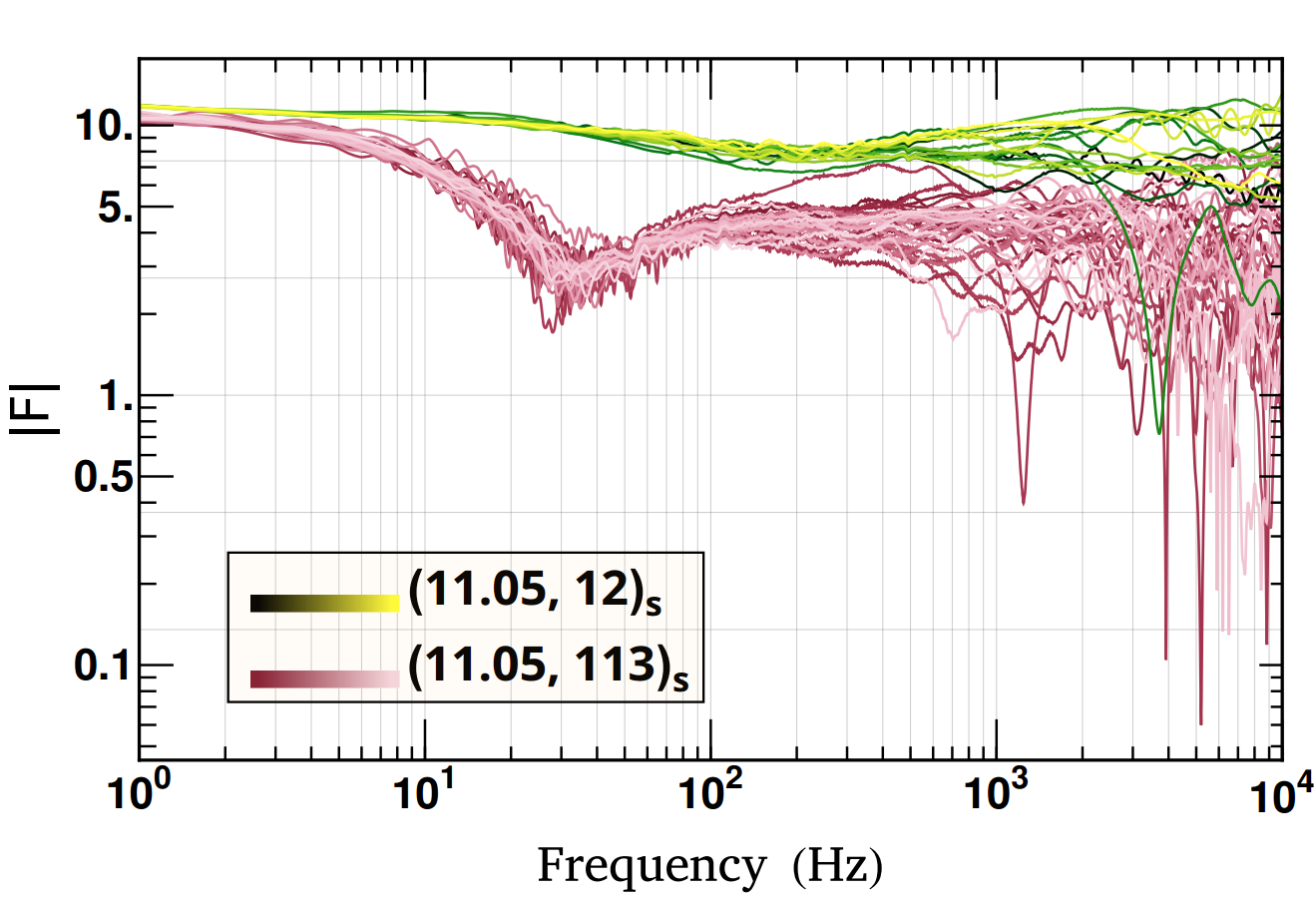}
    \caption{
    Effect of varying surface microlens density in case of \corr{a saddle point} macroimage near critical curve ($\sqrt\mu\sim11.05$ in this case). $|F|$ represents the absolute value of the amplification factor and the comparison is shown between two densities, $\Sigma_\bullet\in\{12, 113\}$. We show 36 realisations for the case $(11.05, 113)_\mathrm{s}$ and 16 realisations for $(11.05, 12)_\mathrm{s}$.}
    \label{fig:|F|_sad_amp11_sd12_sd113)}
\end{figure}
\begin{figure*}
    \centering
    \includegraphics[scale=0.322]{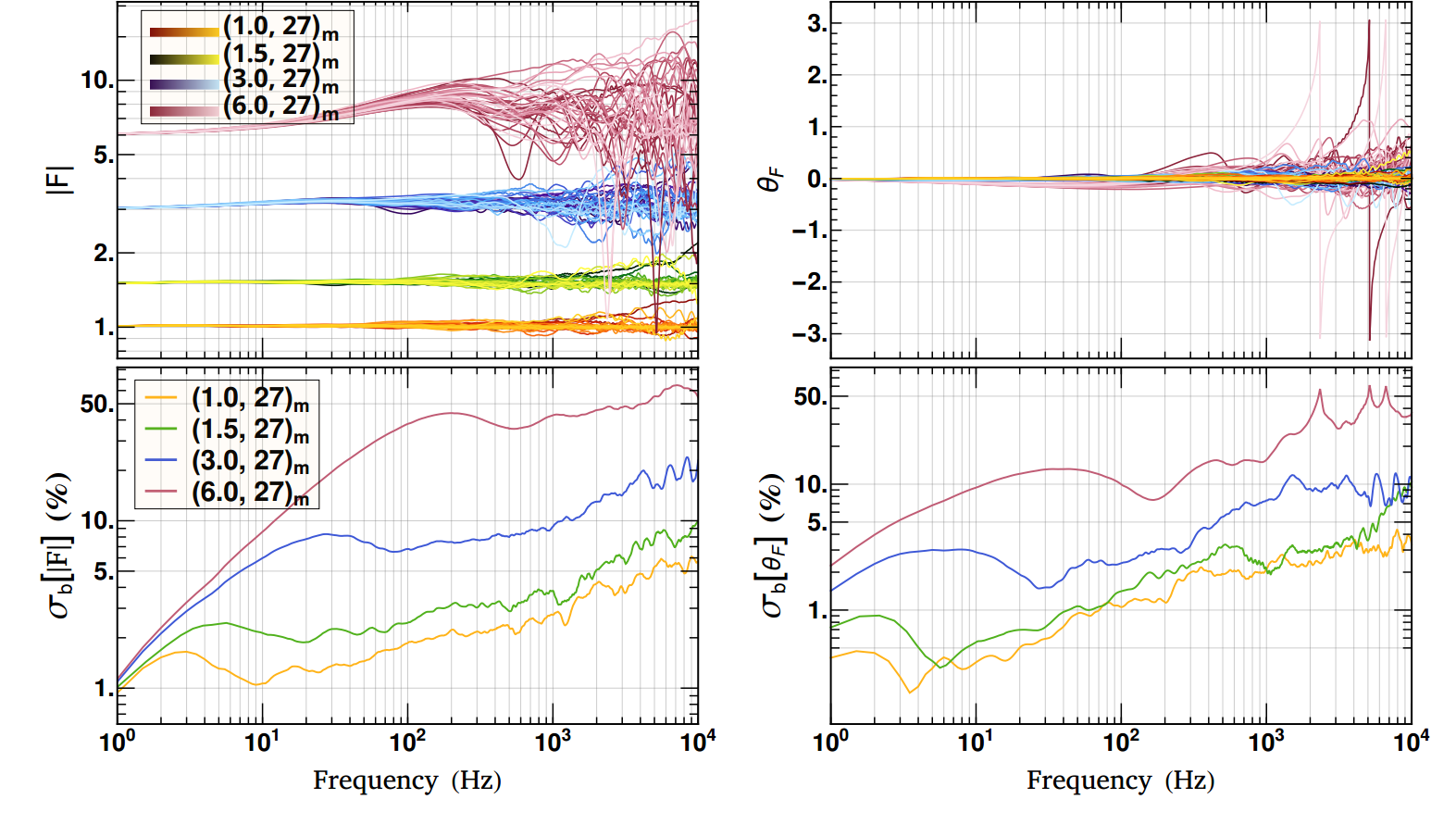}
    \caption{Effect of varying macro-amplification ($\sqrt{\mu}$). We show the comparison between the amplification 
    curves corresponding to $\sqrt{\mu}\in \{1.0,~1.5,~3.0,~6.0\} $ for a fixed surface microlens density 
    of $\Sigma_\bullet=27$ M$_\odot$~pc$^2$. The top row shows the amplification factor, $|F|$, and the phase shift,
    $\theta_{\rm F}$, of 36 realisations for each of the four cases. The bottom row shows the 
    standard deviation of realisations relative to the strong lens limit 
    ($|F|=\sqrt{\mu}$ and $\theta_F=0$) as a function of the frequency (see Eqs. \ref{eq:sigma_sdv_amp_imf} and \ref{eq:sdv_G_b}).}
    \label{fig:F(f)_amp_variation}
\end{figure*}

The colours of the markers help us study the effect of different types of population in our realisation. It is 
worth noting that the realisations corresponding to the extreme behaviours are not necessarily the ones containing
a heavy star. 
This inference shows that in a typical case of GW lensing, low mass microlenses with $\mathcal{O}(1)\ M_\odot$ cannot be neglected in general, especially when they are closer to the source and when \corr{the} macro-magnification value is significantly high, which increases the effective mass of the microlenses \citep[e.g., ][]{2018ApJ...857...25D}. 
This implies that the distribution of microlenses around the source is a crucial factor for studying microlensing due to a population and we discuss this further in \Sref{ssec:ml_effect_pmdist}. It also suggests that the inverse problem of finding the microlensing configuration from lensed GWs is a non-trivial problem, in general. 

In \Fref{fig:|F|_sad_amp11_sd12_sd113)}, we show the effect of varying the density near \corr{a saddle point} macroimage for a high macro-amplification value of $11.05$. The densities considered are $113~$M$_\odot$pc$^{-2}$ and $12~$M$_\odot$pc$^{-2}$ where the latter correspond to densities in the intracluster regime. This comparison suggests that the strong modulations and overall deamplification in case of $(11.05,113)_\mathrm{s}$ is mainly due to the density of the microlens population in addition to the high macro-amplification value, since we do not observe such strong behaviour when we lower the density, as shown in the case of $(11.05,12)_\mathrm{s}$. 
 
\corr{Therefore, the presence of microlenses near saddle point macroimages gives a relatively more significant effect at lower frequencies, as also pointed out in the previous subsection \ref{ssec:ml_effect_saddle}}.
This is because the microcaustics around \corr{saddle-type} macroimages produced by the microlenses are found to have larger regions of low magnifications than those found around \corr{minima-type} macroimages (see e.g., \citealt{2018ApJ...857...25D}; see figures 5 and 10 in \citealt{2019A&A...627A.130D}). Thus, the probability that the microlenses will cause deamplification of the signal is much higher for the \corr{saddle-type} macroimages consistent with what we find.

\subsection[Effect of varying macro-amplification value]{Effect of varying macro-amplification ($\sqrt{\mu}$) value}
\label{ssec:ml_effect_vary_amp}
In this subsection, we discuss the effect of varying macro-amplification, $\sqrt{\mu}$, on the 
amplification factor $F(f)$. The surface microlens density has been fixed to $\Sigma_\bullet=27$~M$_\odot$ pc$^{-2}$ while $\sqrt{\mu}\in \{1.0,~1.5,~3.0,~6.0\}$. 
The comparison is done by drawing 36 realisations for each of the four sets of $(\sqrt{\mu},\Sigma_\bullet)_\mathrm{m}$. 
The results are shown in \Fref{fig:F(f)_amp_variation}.

In the top panel of \Fref{fig:F(f)_amp_variation}, we clearly observe that the overall scatter among the 
realisations increases significantly as we increase $\sqrt{\mu}$, which implies that microlensing effects are 
strongly dependent upon $\sqrt{\mu}$. 
For example, for the realisations corresponding to $\sqrt{\mu}\in \{1.0,\ 1.5\}$, the microlensing effects are not much significant (i.e., $|F|\sim\sqrt{\mu}$ and $\theta_F\sim0$) as compared to those for higher $\sqrt{\mu}$ values, especially, at lower frequencies ($f\lesssim10^3~$Hz).  
\corr{This becomes even more obvious when we compute the standard deviation of the realisations after subduing the effect of strong lensing, which can be defined as}
\corr{
\begin{equation}
\sigma_{\text{\scriptsize (b, c)}}^2=\frac{1}{N}\sum\limits_{i}^{N=36} G^{\text{\scriptsize (b, c)}}_i, 
\label{eq:sigma_sdv_amp_imf}
\end{equation}
}
\corr{such that
\begin{equation}
       G^{\text{\scriptsize b}}_i= 
\begin{cases}
    \left[\frac{|F_i|}{\sqrt{\mu}}-1\right]^2,& \text{for}\hspace{0.2cm} \sigma_{\text{\scriptsize b}}[|F|]\\[8pt]
   \left[\theta_{F_i}-0\right]^2,   & \text{for}\hspace{0.2cm} \sigma_{\text{\scriptsize b}}[\theta_F]
\end{cases}
\label{eq:sdv_G_b}
\end{equation}
and
\begin{equation}
       G^{\text{\scriptsize c}}_i= 
\begin{cases}
    \left[\frac{|F_i|-|F|_{\rm avg.}}{\sqrt{\mu}}\right]^2,& \text{for}\hspace{0.2cm} \sigma_{\text{\scriptsize c}}[|F|]\\[8pt]
   \left[\theta_{F_i}-\theta_{F \rm{avg.}}\right]^2,   & \text{for}\hspace{0.2cm} \sigma_{\text{\scriptsize c}}[\theta_F].
\end{cases}
\label{eq:sdv_G_c}
\end{equation}
Here, $\sigma_{\rm b}$ and $\sigma_{\rm c}$ are frequency-dependent where the former is relative to the strong lensing limit and the latter relative to the average value of the realisations. 
The $\sigma_{\text{\scriptsize c}}$ is used in the next section.
The notations $[|F|]$ and $[\theta_F]$ are used to indicate the standard deviations associated with the respective quantities.
The standard deviations thus computed are purely dependent upon the microlensing effects 
and rise with increasing frequency which signifies that, overall, microlensing effects rise as one moves towards higher frequencies 
(see the bottom panel of \Fref{fig:F(f)_amp_variation}). }
Furthermore, as we vary $\sqrt{\mu}$ from $1.0$ to $6.0$ and analyse $F(f)$ at high frequencies, we observe that $\sigma_{\text{\scriptsize b}}[|F|]$ rises from $\sim 6\%$ to $60\%$, while $\sigma_{\text{\scriptsize b}}[\theta_F]$ shows an increase from $\sim~4\%$ to $40\%$. Such a steep rise clearly demonstrates that the microlensing effects are strongly
dependent upon the macro-amplification value $\sqrt{\mu}$.

Furthermore, from the bottom panel of \Fref{fig:F(f)_amp_variation}, one can observe that the first local maximum (convex-like bump) shifts to higher frequencies with increasing $\sqrt{\mu}$ values accompanied by a significant rise in its magnitude. The same behaviour can be noticed in the individual realisations corresponding to different $\sqrt{\mu}$ values.
This is because, generally, the small-scale microlensing features are negligible at lower frequencies (higher time-delays). However, as the macro-magnification value increases, the amplification of the microimages also increases and becomes significant, leading to a large-scale coherent rise in the microlensing effects. 
As a result, the strength of this coherent rise in $F(f)$ and the frequency up to which the behaviour is coherent will be proportional to the value of $\sqrt{\mu}$. 
Similarly, in \corr{saddle points}, we notice similar but contrary behaviour, i.e., instead of a local maximum (bump), we observe a local minimum (dip) being affected in the same manner at lower frequencies (see \corr{$F(f)$ plots} in \Fref{fig:F(f) saddle}). 
This overall feature can be better understood by analysing the $\widetilde{F}(t)$ curves. For example, \corr{in \Fref{fig:F(f) minima},} one can examine the $\widetilde{F}(t)$ curves corresponding to the $F(f)$ curves and notice the behaviour at time delays $\sim 10^{-2}-10^{-1}$ s. 
Except for the case of extreme magnification ($10.01,~108$), we can see how strong lensing dominates that region resulting in a coherent behaviour. 
The case ($10.01,~108$) is an exception because there we get two very dominant microimages which dominate the overall microlensing effect, unlike in other cases.   
\begin{figure*}
    \centering
    \includegraphics[scale=0.313]{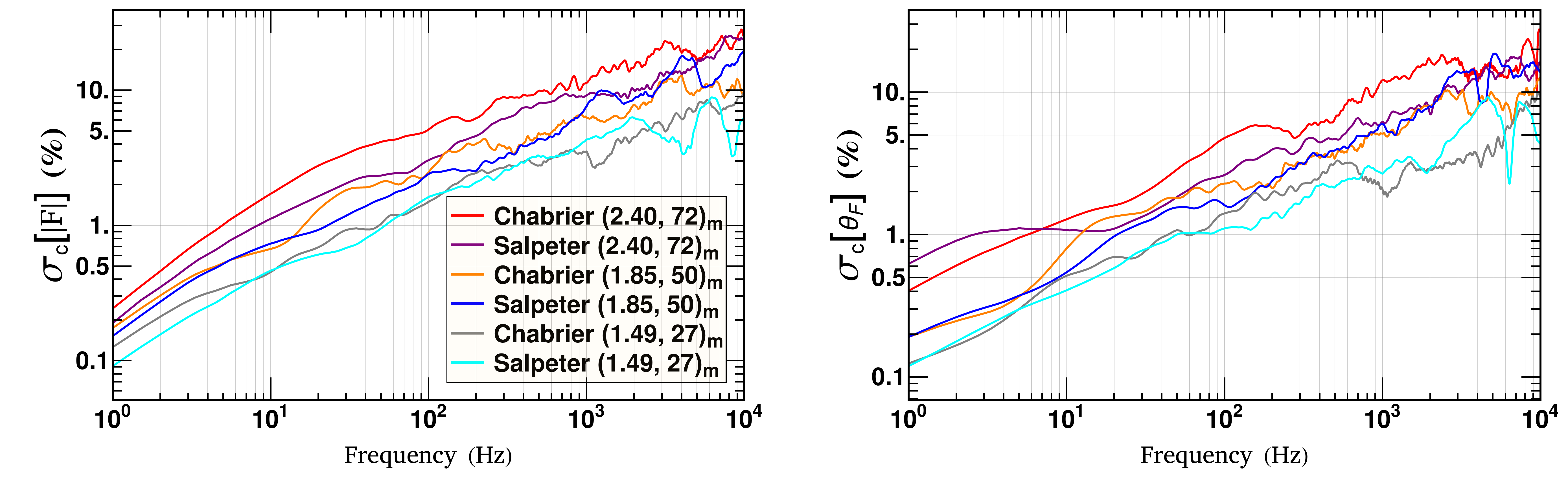}
    \caption{\corr{
    Effect of different stellar IMFs. The comparison 
    is shown for three cases, $(\sqrt{\mu},\Sigma_\bullet)_i\in \{(1.49,27)_{\rm m}, (1.85,50)_{\rm m},(2.40,72)_{\rm m}\}$. 
    For each case, we show 
    the scatter among the realisations in the absolute value $|F|$ (left) and the phase shift $\theta_{\rm F}$ (right), respectively, of the amplification factor -- as defined in Eqs. \ref{eq:sigma_sdv_amp_imf} and \ref{eq:sdv_G_c}.} 
    }
    \label{fig:imf_comparison}
\end{figure*}

\subsection{Effect of varying Stellar Initial Mass Function}
\label{ssec:ml_effect_imf}
Typically, lensing galaxies are of early-type with old stellar population. Knowledge of the stellar 
Initial Mass Function (IMF) is an important pursuit to understand galaxy formation and evolution. 
Many studies aim to understand if there is a universal IMF that can describe all early-type galaxies 
and if it evolves as a function of redshift~\citep[e.g.,][]{2010ARA&A..48..339B, 2020ARA&A..58..577S}. 
Thus far, there is no consensus on a single universal form for the IMF. The 
Chabrier~\citep{2003PASP..115..763C} and Salpeter~\citep{1955ApJ...121..161S} IMFs are some of the 
standard IMFs known to fit early-type galaxies well~\citep[e.g.,][]{2017ApJ...846..166L, 2018MNRAS.479.2443V} 
including those from strong lensing 
studies~\citep[e.g.,][]{2010ApJ...709.1195T, 2010MNRAS.409L..30F, 2015MNRAS.449.3441S, 2016MNRAS.459.3677L, 2019A&A...630A..71S}. 

To investigate the effects of stellar IMF on the amplification, we compare Salpeter IMF 
with our default choice of IMF in this work, i.e., Chabrier. The comparison is done at minima by drawing 36 realisations for three cases of $(\sqrt{\mu},\Sigma_\bullet)_i \in \{(1.49, 27)_\mathrm{m},~(1.85, 50)_\mathrm{m},~(2.40, 72)_\mathrm{m}\}$.
The choice of these numbers is based on the fact that the corresponding strong
lensing magnification ($\mu$) values cover the range of magnification that 
we find in typical strong lens systems. The results are shown in \Fref{fig:imf_comparison}.

\corr{ \Fref{fig:imf_comparison} shows the standard deviation $\sigma_{\text{\scriptsize c}}$ among the realisations as a function of frequency. $\sigma_{\text{\scriptsize c}}$ is defined in \Eref{eq:sigma_sdv_amp_imf} where $G_i^{\text{\scriptsize c}}(f)$ are given in \Eref{eq:sdv_G_c} and the summation is over all $36$ realisations.
The quantities avg.[$|F_i(f)|$] and avg.[$\theta_{F_i(f)}$] denote 
the average value of $F(f)$ obtained among our realisations at a frequency $f$. 
The standard deviations thus computed are purely dependent upon the microlensing effects.
On closer inspection, it is evident from the plots that the effect of microlensing in the case 
of Salpeter IMF becomes prominent at a relatively higher frequency than that of Chabrier IMF. 
For example, observing the lower frequency 
range $\lesssim 200$ Hz, we find that the scatter in the case of Chabrier IMF is more. 
Such behaviour is expected because the Salpeter IMF is comparatively more bottom-heavy than the Chabrier IMF, i.e., it contains a significantly high number of low-mass microlenses compared to the Chabrier IMF. 
Owing to this effect, in the case of the Salpeter IMF the time delay of microimages (relative to macrominima) with significantly high amplification is, on average, relatively small, which is why distortions come later in $F(f)$ for the Salpeter IMF.} 

Since the differences are not significant in the LIGO frequency band, possible constraints on 
the IMF from strongly lensed GWs will require further investigation.

\subsection{Effect of Microlens Population: Mismatch Analysis}
\label{ssec:ml_effect_mismatch}
Here, we quantify the effects of microlensing in the case of strongly lensed GW signals from 
compact binary coalescences (CBCs) by computing the \textit{match} (noise-weighted normalised 
inner product) between the unlensed and the corresponding lensed waveforms, \corr{$m(h_\text{unlensed}, h_\text{lensed})$ 
(see \Aref{appendix: mismatch_details}). 
Mismatch, $\mathcal{M}$, and the match, $m$, are related by a simple relation $\mathcal{M}=1-m$ (see \Eref{Eq:match_mismatch}).}
For generating the unlensed waveforms and computing 
the match, we have used \corr{the} \texttt{PyCBC} package (\citealt{alex_nitz_2020_4355793}, 
\citealt{ 2016CQGra..33u5004U})\corr{, and worked with the approximant \texttt{IMRPhenomPv3} and an $f_\text{\scriptsize low}$ value of $20$ Hz, where $f_\text{\scriptsize low}$ is the starting frequency of the GW waveform}. The lensed GW waveforms have been obtained by modifying the 
unlensed waveforms using the realisations of the amplification factor $F(f)$ for $(\sqrt{\mu},\Sigma_\bullet)_i$ 
as shown in Figs. \ref{fig:F(f) minima} and \ref{fig:F(f) saddle}. Due to normalisation, \corr{the} match 
does not contain any information about the distance of the source or the strong lensing amplification 
value. Therefore, it is a pure measure of the microlensing effects.
We first discuss the results for some extreme cases and then for the typical cases. 
\corr{We only focus on the waveforms corresponding to non-spinning and non-eccentric binaries in this paper.}
For the case of $(\sqrt{\mu},\Sigma_\bullet)_\mathrm{s}$ 
corresponding to saddle points, we have applied a low pass filter to the realisations before computing 
the match in order to remove the small-scale oscillations introduced due to numerical error and, 
therefore, to not overestimate our mismatch results.

\corr{Generally, a GW waveform from chirping binaries can be classified into three regions: inspiral, merger, and ringdown (in the order they appear in the time and frequency series). 
The inspiral and the merger phase are roughly separated by the GW frequency at maximum GW strain amplitude, $f_\text{\scriptsize ISCO}$,  which is about twice the orbital frequency at the ISCO (inner-most stable circular orbit; \citealt{ligo2017basic}). 
Similarly, the merger and the ringdown (post-merger) phase are roughly separated by the QNM frequency of the real part of the fundamental ($n=0$) $l=m=2$ mode, $f_\text{\scriptsize RD}$, of the remnant Black Hole (e.g., see right panel of \Fref{fig:ML_effect_WF}; \citealt{2006PhRvD..73f4030B}).  
Also, the chirp evolution of the signal is related to the binary masses as  $f(t)\approxprop q^{-3/8}M^{-5/8}$ in the inspiral phase, where M is the total mass of binary and q is the mass ratio such that $q\in(0,1]$ (\citealt{ligo2017basic}).} 

\corr{Because microlensing effects generally increase with increasing frequency, one should expect mismatch to also increase as the GW signal's length associated with high-frequencies increases. Consequently, the behaviour of $F(f)$ near $f_\text{\scriptsize ISCO}$ and in the merger phase becomes essential. 
However, the maximum power of the GWs is usually contained in the inspiral phase, followed by the merger and the ringdown phases. 
Therefore, the nature of $F(f)$ in the inspiral phase (at low frequencies), where microlensing effects are not usually significant, becomes crucial in estimating the overall mismatch. 
Unfortunately, in the ringdown phase, which lies in the regime where microlensing shows interesting behaviour, the signal's length, and power are relatively small, and the signal is buried deep in the noise. 
Due to this, the current ground-based detectors have not yet been able even to extract signals present in this regime. 
As a result, even though the ringdown phase is significantly affected, it does not contribute notably to the overall mismatch. 
So, for a given GW signal, since microlensing effects usually increase with increasing frequency, the mismatch will depend mainly upon the signal's length associated with the high-frequency band $(f_\text{\scriptsize ISCO} \lesssim f \lesssim f_\text{\scriptsize RD})$ and the nature of $F(f)$ in that regime. However, when the modulations in $F(f)$ are significant even at low frequencies $(f_\text{\scriptsize low} \lesssim f \lesssim f_\text{\scriptsize ISCO})$, a waveform with gradual chirp evolution will also develop a higher mismatch.  
This knowledge helps us in predicting and analysing the behaviour of the mismatch introduced due to microlensing.
}

\begin{figure*}
    \centering
    \includegraphics[scale=0.28]{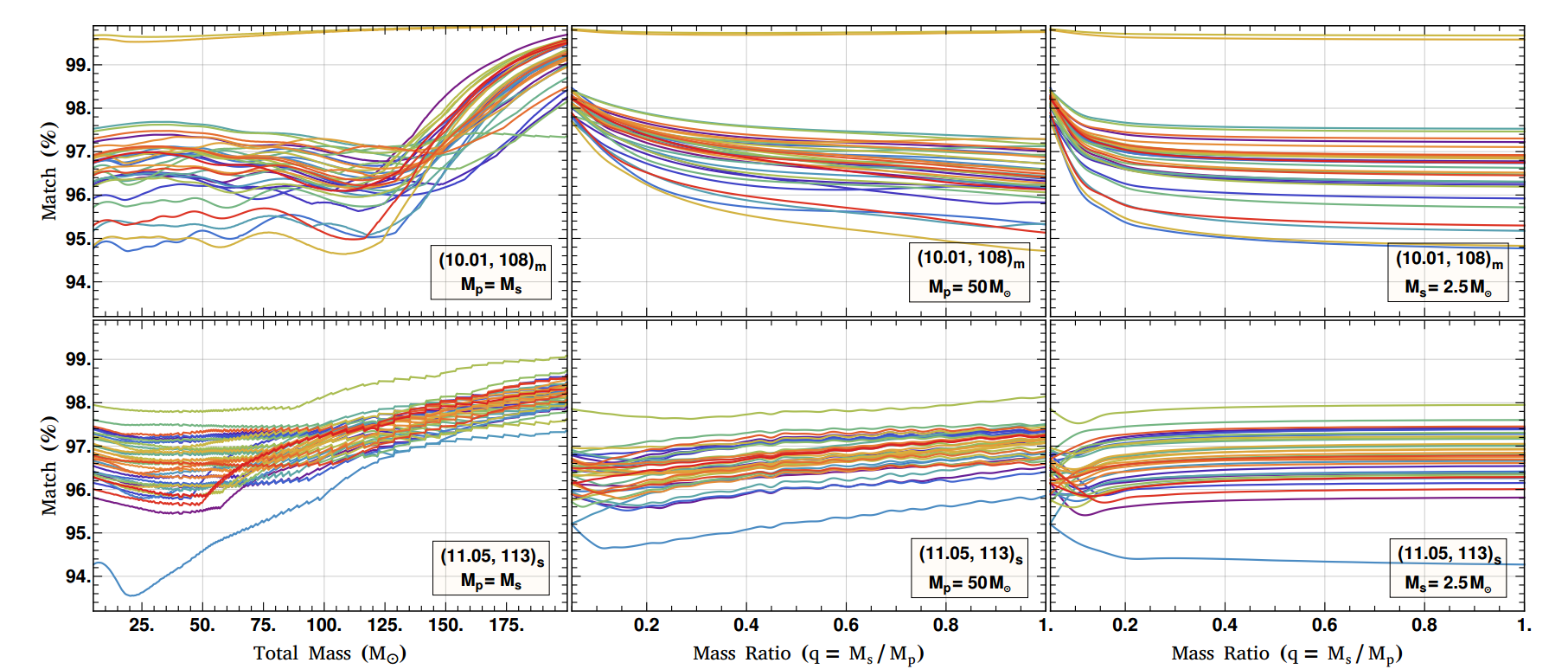}
    \caption{
    Match between the unlensed and the corresponding lensed waveforms for highly magnified minima (top row) 
    and saddle point (bottom row) macroimages. The lensed GW waveforms have been obtained by modifying 
    the unlensed waveforms using the respective realisations of the amplification factor $F(f)$ as shown 
    in Figs. \ref{fig:F(f) minima} and \ref{fig:F(f) saddle} (keeping the same colouring scheme). In all 
    the plots, $M_{\rm p}$ and $M_{\rm s}$ represent the primary and secondary mass of a binary, 
    respectively, such that $M_{\rm p}\geq M_{\rm s}$ and mass ratio $q=M_{\rm s}/M_{\rm p}\in (0,1]$. 
    The first column shows match as a function of the total mass of the (non-spinning) binaries. The second 
    column shows match as a function of the mass ratio $q$ of the binary for a fixed value of $M_{\rm p}=50$ 
    M$_\odot$, while the third column shows the same for a fixed value of $M_{\rm s}=2.5$ M$_\odot$. }
    \label{fig:mismatch_plots}
\end{figure*}
\begin{figure*}
    \centering
    \includegraphics[scale=0.5]{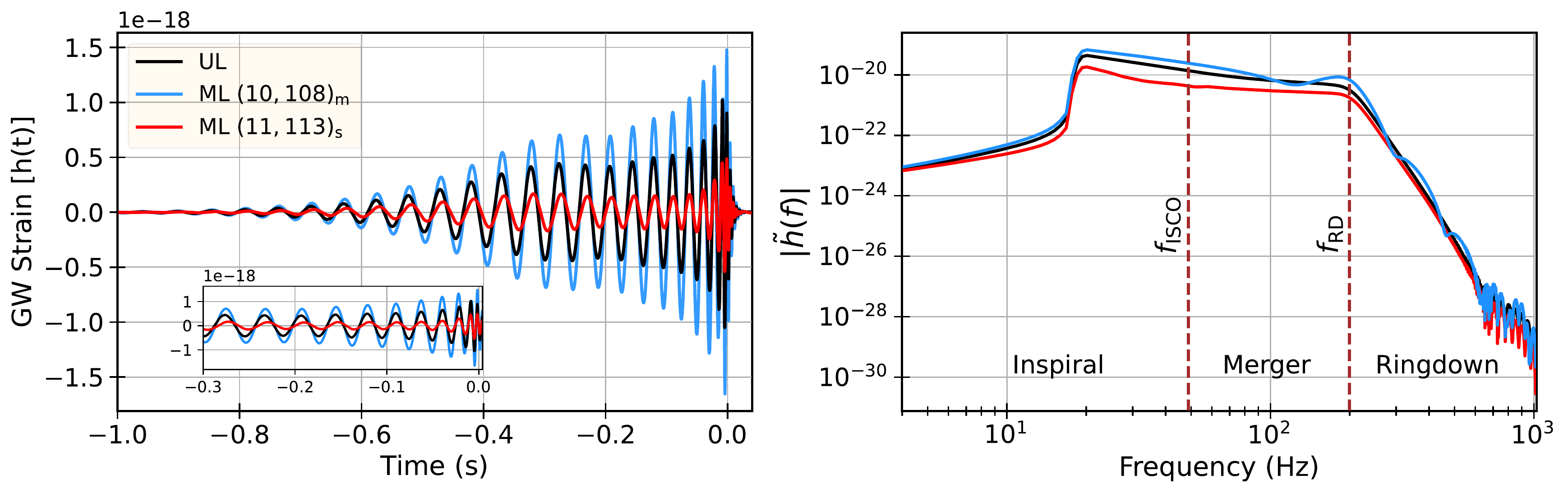}
    \caption{\corr{
    Effect of microlensing on a GW waveform of 45~M$_\odot$+45~M$_\odot$=90~M$_\odot$ binary. We show the GW strain amplitude, $h(t)$, as a function of time (left) and the absolute value of its Fourier transform, $|\tilde{h}(f)|$, as a function of frequency (right). With the unlensed waveform as a reference (black), we show the microlensing effect for two of the realisations (blue and red) from our simulations which display extreme behaviour. These realisations correspond to $(\sqrt{\mu},\Sigma_\bullet)_i\in \{(10.01, 108)_{\rm m}, (11.05, 113)_{\rm s}\}$ and produce a
    mismatch of $\sim 5.1\%$ and $\sim 4.4\%$, respectively. The effect of strong lensing has been subdued, i.e., $F(f)\rightarrow F(f)/\sqrt{\mu}$. We also show the three phases of GW emission on the right panel categorised on the basis of $f_\text{\scriptsize ISCO}$ of the merging binaries and $f_\text{\scriptsize RD}$ of the remnant black hole.
    }}
    \label{fig:ML_effect_WF}
\end{figure*}

\begin{figure}
    \centering
    \includegraphics[scale=0.29]{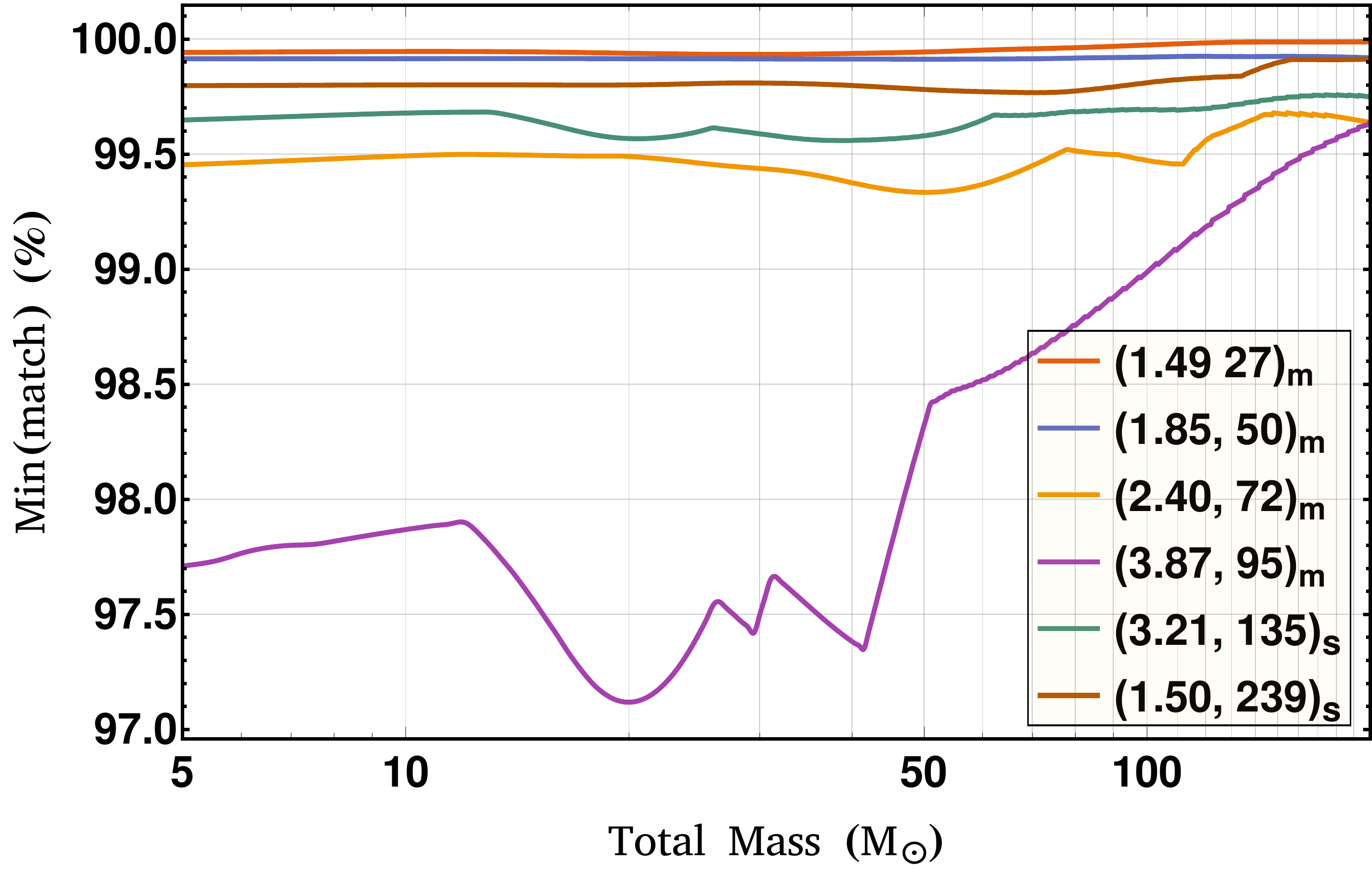}
    \caption{Minimum match (or maximum mismatch) among realisations (shown in Figs. \ref{fig:F(f) minima} and 
    \ref{fig:F(f) saddle}),  corresponding to typical values of $(\sqrt{\mu},\Sigma_\bullet)_i$, as we vary the total mass 
    of the binary.  The result is based upon 36 realisations of $F(f)$ computed for each of the six cases. The result for extreme cases of $\sqrt{\mu}$ is shown in \Fref{fig:mismatch_plots}.}
    \label{fig:min_msm_typical}
    \vspace{-0.5cm}
\end{figure}

\subsubsection{Extreme cases}

In \Fref{fig:mismatch_plots}, we show the match values for two cases of $(\sqrt{\mu},\Sigma_\bullet)_i$, 
namely, $(10.01, 108)_\mathrm{m}$ and $(11.05, 113)_\mathrm{s}$ (top and bottom panels, respectively). These cases correspond to the extreme ones that we have covered for minima and saddle type images, respectively, as they correspond to a large macro-magnification value ($\sim 10^2$) compared to the typical values ($\mu \lesssim 10$, see \Tref{tab:stellar_density}). The primary ($\rm M_p$) and the secondary ($\rm M_s$) mass of a binary are chosen such that $\rm M_{\rm p}\geq M_{\rm s}$ and mass ratio 
$q=\rm M_s/M_p\in (0,1]$. 
\corr{We use this definition of mass ratio throughout our analysis.}
The first column shows the match as a function of the total mass of the (non-spinning) binary for $q=1$, while the second and the third columns show the match as a function of the mass ratio parameter for $\rm M_p=50$ M$_\odot$ and $\rm M_s=2.5$ M$_\odot$, respectively.

\corr{We notice that, in extreme cases, the mismatch values are high and can even exceed $\sim 5\%$. As a general trend, we observe that for $q=1$, on average, the mismatch increases as we lower the mass of the binary (first column). 
This trend is observed because the smaller mass binaries have a relatively higher power in the high frequencies, and they have relatively higher values of $f_\text{\scriptsize ISCO}$ and $f_\text{\scriptsize RD}$, where microlensing effects are relatively stronger. However, since they also have longer GW waveforms, if the distortions in $F(f)$ are not as significant at low frequencies (which is usually the case), then the mismatch may as well decrease due to the presence of a greater length of the signal unaffected by the lens.  
Due to these two competing effects, we do not see a monotonic rise in the match initially as we increase the binary mass in the case of minima (first row, first column). Whereas for the saddle case considered (first row, second column), we do observe a coherent rise because of stronger modulations of $F(f)$ even in the low-frequency regime (see \Fref{fig:F(f) saddle}), causing a significant mismatch even in the inspiral phase. 
In the case of GWs from chirping binaries with unequal component masses, the length of the signal increases with decreasing mass ratios when $M_{\rm p}$ is held fixed and with increasing mass ratios when $M_{\rm s}$ is held fixed. We can then apply the same argument for the second and third columns as in the first column and explain the behaviour because of the trade-off between the two opposing effects.}

\corr{Lastly, in \Fref{fig:ML_effect_WF}, we explicitly show how microlensing in such extreme cases can affect a GW waveform. 
The waveform associated with an equal mass binary ($q=1$) of $90$M$_\odot$ is microlensed by considering two realisations of $F(f)$ that gave maximum mismatch for each $(\sqrt{\mu},\Sigma_\bullet)_i\in \{(11.05, 113)_{\rm s}, (10.01, 108)_{\rm m}\}$. 
The mismatch obtained for the chosen realisations corresponding to the minimum and the saddle point are $\sim 5.1\%$ and $\sim 4.4\%$, respectively.
The modified waveforms due to these realisations are computed using \Eref{Eq:F(f)_effect_fdwf} after removing the effect of strong lensing, i.e., after scaling $F(f)\rightarrow F(f)/\sqrt{\mu}$.
The left panel shows the effect on the time domain waveform, while the right panel shows it in the frequency domain. 
In the left panel, we can see the modulations in both the amplitude and the phase of the GW strain signal $h(t)$ (see the inset for better visibility). 
In the right panel, we plot the absolute value of the frequency-domain waveform $|\tilde{h}(f)|$, where $\tilde{h}(f)$ is the Fourier transform of the timeseries $h(t)$. 
Consequently, the ordinate can also be interpreted as the power spectrum of the signal.} 

\corr{The vertical dashed maroon lines, we show $f_\text{\scriptsize ISCO}\approx 48~\rm Hz$ and $f_\text{\scriptsize RD}\approx 200~\rm Hz$ for this case (\citealt{1972ApJ...178..347B}, \citealt{2006PhRvD..73f4030B}, \citealt{2008PhRvD..78h1501T}), separating the three phases of the GW emission: inspiral, merger, and the ringdown.
As previously mentioned, one can observe that most of the power is contained in the lower frequencies ($f<f_\text{\scriptsize ISCO}$) as it constitutes the inspiral phase of merging binaries. 
We also see interesting modulations here. For the realisation corresponding to the saddle point (red curve), we mainly observe modulations in the lower frequencies, whereas for the realisation corresponding to the minimum (blue curve), we mainly observe modulations at higher frequencies.
This distinction is because of the different behaviour of $F(f)$ in the two extreme cases of our simulation (see the last row of \Fref{fig:F(f) minima} and the first row of \Fref{fig:F(f) saddle}).
The modulations of the blue curve in the ringdown phase are particularly of great interest, as they can be a discerning feature of microlensing. 
However, since these post-merger signals are challenging to extract, such searches for microlensing features can only be done once we have interferometers with high enough sensitivity, especially at high frequencies. 
}
\begin{figure*}
    \centering
    \includegraphics[scale=0.209]{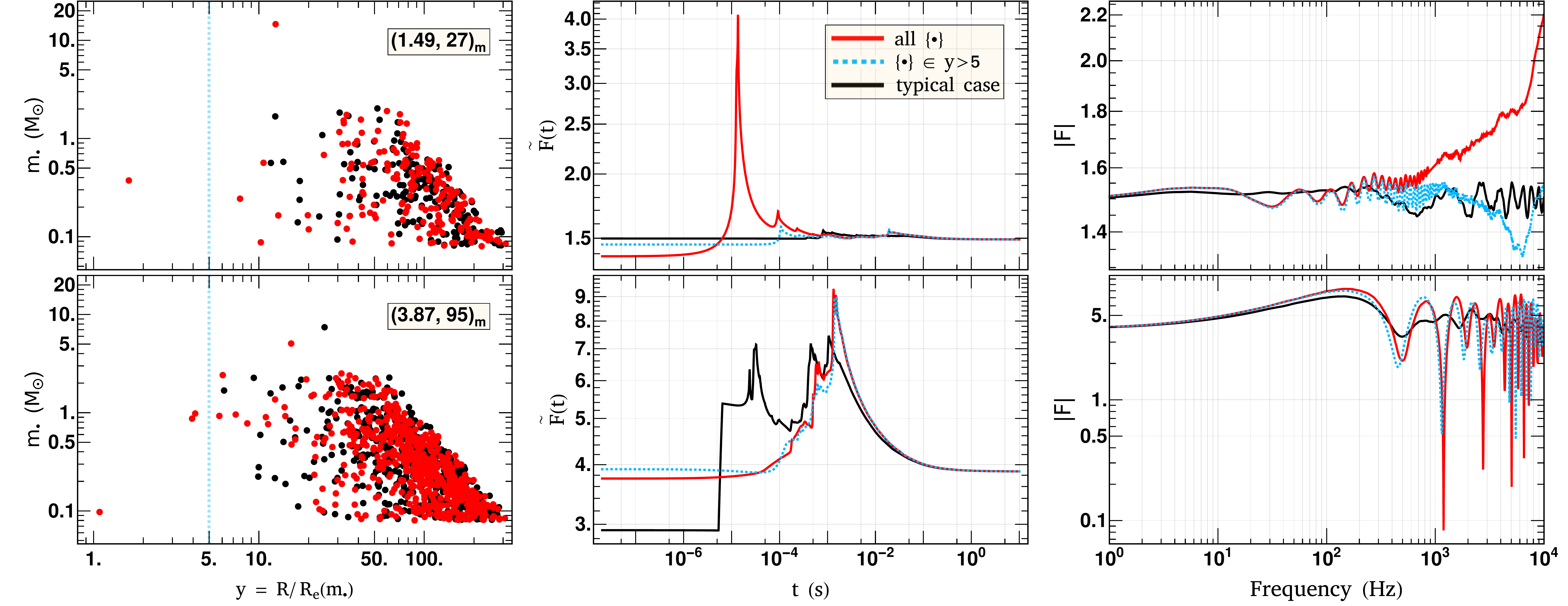}
    \caption{
    {\it Left}: Microlens mass ($m_\bullet$) as a function of their radial distribution with respect to the macroimage of a source. The impact parameter $y$ is in the units of $R_{\rm e}(m_{\bullet})$, the Einstein radius for a microlens of mass $m_{\bullet}$. 
    The distribution in red and black corresponds to a chosen atypical and typical case \corr{of $F(f)$} out of 36 realisations generated for a specific $(\sqrt{\mu},\Sigma_\bullet)_i$ (see labels in the top and bottom panels), as shown in \Fref{fig:F(f) minima}.
    {\it Middle}: The $\widetilde{F}(t)$ as a function of the time delay with respect to the macro$-$minima. The red and black corresponds to the same population of atypical and typical cases, respectively, as shown in the left column. Additionally, the blue curve corresponds to the atypical such that the effect from all of the microlenses with impact parameter $y<5$ is excluded. Comparison of red and blue curves highlights the contribution of microlenses at low impact parameters. {\it Right}: The modulus of the amplification factors, $|F(f)|$,  obtained from the corresponding $\widetilde{F}(t)$ curves in the middle column. }
    \label{fig:m_y_pop_analysis}
\end{figure*}

\subsubsection{Typical Cases}

We calculate the minimum match among the 36 realisations for each case of $(\sqrt{\mu},\Sigma_\bullet)$ which correspond to the typical conditions found in lensing galaxies for both minima and saddle points (see \Fref{fig:min_msm_typical}). The minimum match is calculated as a function of the total mass of the binary.

It is worth noting that for typical cases, the mismatch hardly exceeds one percent and is 
mostly within sub-percent levels. The case $(3.87, 95)_\mathrm{m}$ can be considered to lie within typical and extreme cases (as $\mu\sim15$). In such cases, mismatch can \corr{even exceed} $\sim 2.5\%$ depending upon the source parameters.
Therefore, while  microlensing will not affect the detectability of GW signals in typical situations, 
it might still affect the estimation of the source parameters in a significant way, as its influence on the waveform may be degenerate with the GW source parameters \corr{e.g., mass and spin of the binary components.}


\subsection{Effect of Microlens Population: Dependency on Microlens Distribution around the Source}
\label{ssec:ml_effect_pmdist}

\corr{In this subsection, we attempt to trace the origin of microlensing imprints on the amplification factor $F(f)$ by analysing the mass and the spatial distributions of the microlens around a macroimage of the source. Although we only show a few selected cases here, our conclusions are based upon a thorough analysis of the realisations in Figs. \ref{fig:F(f) minima} and \ref{fig:F(f) saddle}.}

In \Fref{fig:m_y_pop_analysis}, we study the effect of the distribution by analysing a pair of typical and atypical realisations of \corr{$F(f)$}, generated for \corr{$(\sqrt{\mu},\Sigma_\bullet)_i\ \in\ \{(1.49,~27)_{\rm m},(3.87,~95)_{\rm m}\}$} (see  \Fref{fig:F(f) minima}).  
By atypical cases, we mean the realisations that produced maximum mismatch (minimum match) in \Fref{fig:min_msm_typical}, which can also be determined by visual inspection in \Fref{fig:F(f) minima}. 
In the leftmost column, we plot the microlens mass ($m_{\bullet}$) as a function of the dimensionless impact parameter ($y$). \corr{ Here, the usage of the symbol `$y$' is the same as in the case of an isolated point lens (see \Eref{eq:amp fac point}) as we want to study the effect of each microlens in terms of its impact parameter. Therefore, we can write} $y(m_{\bullet})=R/R_{\rm e}(m_{\bullet})$ such that $R$ is the projected distance between the microlens and the source on the image plane and $R_{\rm e}(m_{\bullet})$ is the Einstein radius of a lens of mass $m_{\bullet}$.
\corr{Also, from \Eref{eq:mag_td_pnt_lens}, one can notice that the magnification of the microimages is purely a function of $y$, thereby making it an important quantity to consider while studying the effect of microlenses.} 
\corr{The importance of mass comes from the fact that the impact parameter $y(m_{\bullet})\propto m_{\bullet}^{-1/2}$, while the time delay $\Delta t_{\rm d}=T_{\rm s}\Delta\tau_{\rm d}\propto m_{\bullet}^{1/2}$.}

In the top row of \Fref{fig:m_y_pop_analysis}, we analyse a pair of typical (black) and atypical (red) realisations for the case of $(1.49, 27)_\mathrm{m}$, whose $|F|$ curves are shown in the right panel. The corresponding $\widetilde{F}(t)$ (middle panel) reveals that there are no significant microimages for the typical case (black), whereas there is one significant microimage (peak at $t\sim10^{-5}$~s) for the atypical case (red). Upon inspection of the spatial distribution of microlenses around the source (left panel), we find that there are usually no microlenses with $y<10$ for typical realisations (black). In contrast, there are usually a few microlenses with $y<5$ for many of the atypical realisations (red). To assess the significance of microlenses having $y<5$, we recomputed the $\widetilde{F}(t)$ and $|F|$ for the atypical case considered by excluding the microlenses present in that region (blue curves in the middle and the right panels, respectively), which corresponds to removing only one microlens of mass $m_{\bullet}\sim 0.37$ M$_\odot$ and $y\sim1.6$ from the population marked in red. Despite the low mass of the removed microlens, we observed a drastic difference between the blue and red curves and noticed that the dominant microimage in the red $\widetilde{F}(t)$ curve disappeared altogether in its absence. We also tested the effect of the rest of the microlenses with $y>5$ on the dominant microimage produced by that single microlens and found that the magnification of this dominant microimage is further enhanced by the rest of the nearby microlenses, making it more prominent compared to additional microimages (peaks seen at larger $t_{\rm d}>10^{-5}$~s).

A similar behaviour was observed in the realisations of \corr{$(\sqrt{\mu},\Sigma_\bullet)_i\in \{(1.85, 50)_{\rm m}, (2.40, 72)_{\rm m}\}$}, where the effect of microlenses at low $y\sim\mathcal{O}(1)$ was observed to be more significant.
However, the presence of microlenses with low $y$ does not always guarantee a significant microlensing effect. The effect from nearby microlens population also needs to be considered, which may either enhance its microlensing effect or even suppress it in some cases. We observed that in a few of the realisations where at least one microlens had $y<5$, the microlensing effect was not much significant. Also, in the case of $(1.49, 27)_\mathrm{m}$, there was one realisation corresponding to the atypical case where no microlens was present below $y=5$, but the microlensing effect was still significant in the LIGO range. 

Interestingly, in the bottom row of \Fref{fig:m_y_pop_analysis}, we repeated a similar exercise for the case $(3.87, 95)_\mathrm{m}$ and did not observe such strong dependency on microlenses with low $y$. This observation is clear from the comparison between the red $\widetilde{F}(t)$ curve (middle panel) with the corresponding blue curve, as they do not deviate significantly from each other, unlike in the top row.

Next, we also studied the impact of excluding the microlens population, distributed at larger impact parameters, by applying varying limits on the maximum value of the impact parameter. For instance, we chose \corr{$y_l\equiv y\in\{20, 50\}$} as two possible limits. We then compared the resulting $F(f)$ with the original $F(f)$ limited by our box size. We found that the relative errors between the two were quite low for both $y_l$. The relative errors, on average, were $\lesssim 5\%$ for typical $\sqrt{\mu} \lesssim 2.40$, and had a maximum value of $\sim 10\%$ in the LIGO frequency range. Therefore, one can greatly reduce the computational cost by considering microlenses only up to certain $y_l$ such as $20$ or $50$. However, the relative error in case of $(3.87,~95)_\mathrm{m}$ turned out to be large, and reached around $\sim90\%$ for $y_l=20$ and $\sim 50\%$ for $y_l=50$. The differences seen here for two cases of $(\sqrt{\mu},\Sigma_\bullet)_i$ is consistent with what we observed in top and bottom rows of \Fref{fig:m_y_pop_analysis}. Hence, as we increase $\sqrt{\mu}$ values, more and more microlenses become significant contributors to the microlensing, and one cannot apply such limits for improving the computational efficiency. In fact, we observe that microlenses with very high impact parameters  \corr{$y\gtrsim 100$} can also contribute significantly to the microlensing effects when present near a highly magnified macroimage ($\mu\gtrsim 15$).

\section{Conclusions}
\label{sec:conclusions}

We investigated the effects of microlensing in the LIGO/Virgo frequency band by a population of point mass objects such as stars and stellar remnants embedded in the potential of a macrolens. In particular, we calculated the microlensing effects for various  combinations of surface microlens density and macro$-$magnification, typically found at the location of minima and saddle points in galaxy-scale lenses.
\corr{Additionally, we explored individually the three most crucial macrolens parameters that are responsible for microlensing: macro-magnification, the surface microlens density, and the IMF.}
For each of the investigated cases, we generated 36 realisations to robustly infer general trends and microlensing effects. To understand the impact of microlensing on CBC signals, we computed the match between the unlensed and lensed waveforms. The mismatch thus obtained is a pure measure of microlensing effects as it is independent of \corr{the} macro-magnification value and 
\corr{only weakly depends upon} the extrinsic GW source parameters.
\corr{Lastly, we also studied how the distribution of microlenses around the macroimages of a source can affect the overall microlensing properties and showed its dependency on macro-magnification.}

Our main conclusions from these investigations are as follows.

\begin{itemize}
\item The most important factor for microlensing to be significant is the strong lensing 
amplification value ($\sqrt{\mu}$) regardless of other parameters, such as the stellar density, 
type of images or IMF. 
\corr{This happens due to the fact that the image plane gets compressed by a 
factor of $\mu$ in the source plane leading to high density of overlapping microcaustics.}
Moreover, higher $\mu$  allows relatively 
larger time delays between sufficiently amplified microimages, which causes modulation even at 
lower frequencies (e.g., \citealt{2019A&A...625A..84D}; \citealt{2020PhRvD.101l3512D}). From our 
analyses, we observed that for sufficiently high macrolensing magnifications ($\mu\gtrsim 15$), 
microlensing effects cannot be neglected. 

\item On average, the microlensing population tends to introduce further amplification (de-amplification) 
for minima (saddle points) in the LIGO frequency range. Similar behaviour is also seen in the geometric 
optics limit for minima and saddle points \citep[e.g.,][]{2002ApJ...580..685S,2018MNRAS.478.5081F}.

\item Overall, we observed the microlensing effects to be more pronounced in the case of macroimages at \corr{saddle points}, especially at lower frequencies. The presence of microlenses becomes 
important in the case of \corr{saddle$-$type} macroimages as the probability of the source lying in the region 
of low magnification is significantly higher, unlike in case of \corr{minima$-$type} macroimages 
(e.g., \citealt{2018ApJ...857...25D}; see figures 5 and 10 in \citealt{2019A&A...627A.130D}).

\item With increasing surface microlens density, we find an overall rise in scatter in $F(f)$ and notice that the distortions become significant from relatively lower frequencies. This is because the presence of multiple microlenses at a location can mimic the effect of a heavier microlens, thereby increasing the probability of significantly amplified microimages, with sufficiently high time-delay values, to have interference.

\item The microlens population generated using the Chabrier IMF gives rise to an overall scatter in $F(f)$ larger than the Salpeter IMF, at typical values of $(\sqrt{\mu},\Sigma_\bullet)_i$. 
However, the effect of a bottom-heavy IMF, like Salpeter, increases at higher frequencies and becomes comparable to IMFs like Chabrier, which has a relatively large number of high-mass microlenses. 
Since the impacts of both IMFs on the $F(f)$ are found to be nearly indistinguishable, the exact choice of IMF model will probably not have distinguishable microlensing effects.
\corr{In other words, any microlensing signatures detected in the data may not help constrain the IMF. However, further investigation is needed to confirm this inference and, in general, to study the properties of the intervening microlens population.}

\item For typical $\sqrt{\mu}$, the microlensing effect in the LIGO band arises mainly due to microlenses with low $y$, where $y$ is the impact parameter between the microlens and the source, as measured in units of Einstein radius of that microlens. For instance, even a single microlens of $m_\bullet~\sim~\mathcal{O}(0.1)$~M$_\odot$ with $y\lesssim 5$ can have a significant microlensing effect in the presence of other microlenses. Therefore, we find that the effect of low-mass microlenses  ($m_\bullet\lesssim 1$) cannot be ignored even for strong lenses with typical $\mu$. 
As we increase $\mu$, the size of the critical curves of individual microlenses also increases, due to which microlenses with large $y$ also start contributing. In fact, for high $\mu$ ($\gtrsim 50$), the microlensing effect due to even lower mass microlenses ($m_\bullet\sim \mathcal{O}(0.01)$ M$_\odot$) becomes significant, especially at higher frequencies, when present in abundance. 

Also, for typical macro-magnifications ($\mu\lesssim 6$), one may consider optimising computational speed by limiting microlenses to $y\lesssim 20$. However, such limits will not work for high macro-magnifications, such as $\mu\gtrsim 15$, where microlenses at high impact parameters, \corr{$y\gtrsim 100$}, also start to matter. 
\corr{
\item For a given GW signal, since microlensing effects usually increase with increasing frequency, the overall mismatch between the lensed and the unlensed signal will mainly depend upon the signal's length associated with the high-frequency band $(f_\text{\scriptsize ISCO} \lesssim f \lesssim f_\text{\scriptsize RD})$ and the nature of $F(f)$ in that regime. However, when the modulations in $F(f)$ are significant even at low frequencies $(f_\text{\scriptsize low} \lesssim f \lesssim f_\text{\scriptsize ISCO})$, a waveform with gradual chirp evolution will also develop a higher mismatch. 
}
\corr{
\item We find that in typical cases of $(\sqrt{\mu},\Sigma_\bullet)$, the mismatch between the unlensed and the lensed waveforms is mostly within sub-percent levels ($\lesssim 1\%$). 
Hence, we expect microlensing not to affect the detectability of GW signals (i.e., its \textit{effectualness}) in typical situations. Nevertheless, it may still affect the inferred GW source parameters (i.e., its \textit{faithfulness}) significantly since its effect on the waveform could be degenerate with other GW parameters.
}
\corr{
\item  In extreme cases, we notice that the mismatch values are high and can even exceed $\sim 5\%$. Such high values may leave observable signatures in the waveform and go undetected in the current detectors, especially after the inspiral phase. 
Since GW sources are point-like, when they lie very close to macro-caustics, their magnifications can indeed be very high 
$\mu\sim\mathcal{O}(10^2-10^3)$ and hence, such possibilities cannot be ignored. In fact, our mismatch analysis suggests that the microlensing effects cannot be neglected for $\mu\gtrsim 15$ while inferring the GW source parameters. This will be explored in more detail in the future.
}
 
\end{itemize}

\section*{Acknowledgements}
We would like to thank S. More and V. Prasad for helpful discussions; A. Ganguly and Sudhagar 
S. for their help with the mismatch calculation. We would also like to thank D. Rana, K. Soni and 
S. Banerjee for their help during this work. AKM would like to thank Council of Scientific $\&$ 
Industrial Research (CSIR) for  financial support through research fellowship No. 524007. A. Mishra 
would like to thank the University Grants Commission (UGC), India, for financial support as a 
research fellow.
We gratefully acknowledge the use of high performance computing facilities at IUCAA, Pune.

\section*{Data Availability}
The data underlying this article will be shared on reasonable request to the corresponding authors.

\bibliographystyle{mnras}
\bibliography{bibliography}


\appendix

\section{Quantifying the Effect of Microlensing on GW Waveforms}
\label{appendix: mismatch_details}

In the presence of intervening stellar-mass microlenses, the time delay between the microimages 
can range from $\sim~1-10^3~\mu$s. As a consequence, GW signals from chirping binaries travelling 
through such a region would undergo microlensing (since $ft_{\rm d}\lesssim 1$). This will introduce a 
frequency-dependent modulation, thereby affecting the morphology of the GW signal. 

For a given amplification factor $F(f)$ (defined in \Sref{sec:basic_lensing}), the unlensed GW 
signal, $h_{\rm u}(t)$, and the corresponding lensed waveform, $h_{\rm l}(t)$, are related by the expression
\begin{equation}
    \tilde{h}_{\rm l}(f)=F(f)\tilde{h}_{\rm u}(f).
    \label{Eq:F(f)_effect_fdwf}
\end{equation}
where $\tilde{h}_{\rm l}$ and $\tilde{h}_{\rm u}$ are the Fourier transforms of the timeseries $h_{\rm l}$ and $h_{\rm u}$, 
respectively.
To quantify the effect of microlensing, one can compute the mismatch ($\mathcal{M}$) between the 
unlensed and lensed waveforms \citep{2016CQGra..33u5004U}, defined as 
\begin{equation}
\mathcal{M}=1-{\rm match}(h_{\rm l},~h_{\rm u})=1-\underset{t_0,\phi_0}{\rm max} \frac{\bra{h_{\rm l}}\ket{h_{\rm u}}}{\sqrt{\bra{h_{\rm l}}\ket{h_{\rm l}}\bra{h_{\rm u}}\ket{h_{\rm u}}}},
\label{Eq:match_mismatch}
\end{equation}
where $t_0$ and $\phi_0$ are, respectively, the arrival time and phase of, say, $h_{\rm u}$ and 
$\bra{.}\ket{.}$ is the noise-weighted inner product, defined as
\begin{equation}
    \bra{h_1}\ket{h_2}=4\text{Re }\int^{f_{\rm high}}_{f_{\rm low}} {\rm d}f~\frac{\tilde{h}_1^*(f)\tilde{h}_2(f)}{S_{\rm n}(f)},
\end{equation}
where $h_{\rm i}$ and $\tilde{h}_{\rm i}$ are, respectively, the timeseries signal and its Fourier transform, 
and $S_{\rm n}(f)$ is the single-sided power spectral density (PSD) of the detector noise. 

\bsp	
\label{lastpage}
\end{document}